\documentclass[letter]{aa} %
\usepackage{graphicx}
\usepackage{txfonts}
\usepackage{silence}
\usepackage{hyperref}
\definecolor{myblue}{HTML}{1F77B4}
\definecolor{mygreen}{HTML}{2CA02C}
\definecolor{myred}{HTML}{D62728}
\definecolor{mymagenta}{HTML}{D33682}
\definecolor{codepurple}{HTML}{C42043}

\hypersetup{
unicode=true,           %
colorlinks=true,        %
linkcolor=myred,        %
citecolor=myblue,       %
filecolor=magenta,      %
urlcolor=mygreen        %
}

\usepackage{bookmark}%

\usepackage{graphicx}	%
\usepackage{amsmath}	%

\usepackage{longtable}
\usepackage{booktabs}
\usepackage{xcolor}
\usepackage{xspace}

\newcommand{\brennaninprep}[1]{{\textcolor{myblue}{Brennan. in prep\xspace}}}

\newcommand{\fyq}{SN~2023fyq\xspace}
\newcommand{\ip}{SN~2009ip\xspace}
\newcommand{\bh}{SN~2015bh\xspace}

\newcommand{\halpha}{H$\alpha$\xspace}
\newcommand{\hbeta}{H$\beta$\xspace}

\newcommand{\kms}{~km s$^{-1}$\xspace}
\newcommand{\msun}{M$_{\odot}$\xspace}

\newcommand{\software}[1]{\textsc{#1}}

\usepackage[switch]{lineno}
\begin{document} 

\title{Spectroscopic observations of progenitor activity 100 days before a Type Ibn supernova}

\author{
S. J. Brennan\inst{1} \href{https://orcid.org/0000-0003-1325-6235}{\includegraphics[scale=0.05]{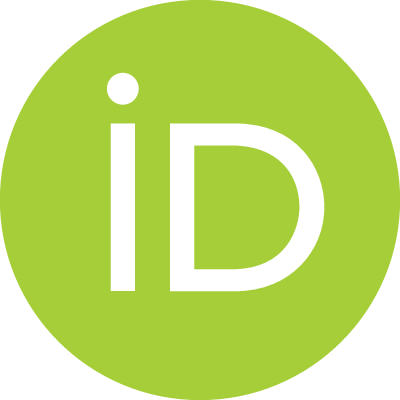}}\and
J. Sollerman\inst{1} \href{https://orcid.org/0000-0003-1546-6615}{\includegraphics[scale=0.05]{figures/ORCIDiD_iconvector.eps}}\and
I. Irani\inst{2} \href{https://orcid.org/0000-0002-7996-8780}{\includegraphics[scale=0.05]{figures/ORCIDiD_iconvector.eps}}\and
S. Schulze\inst{3,4} \href{https://orcid.org/0000-0001-6797-1889}{\includegraphics[scale=0.05]{figures/ORCIDiD_iconvector.eps}}\and
P. Chen\inst{2} \href{https://orcid.org/0000-0003-0853-6427}{\includegraphics[scale=0.05]{figures/ORCIDiD_iconvector.eps}}\and
K. K. Das\inst{5} \href{https://orcid.org/0000-0001-8372-997X}{\includegraphics[scale=0.05]{figures/ORCIDiD_iconvector.eps}}\and
K. De\inst{6} \href{https://orcid.org/0000-0002-8989-0542}{\includegraphics[scale=0.05]{figures/ORCIDiD_iconvector.eps}}\and
C. Fransson\inst{1} \href{https://orcid.org/0000-0001-8532-3594}{\includegraphics[scale=0.05]{figures/ORCIDiD_iconvector.eps}}\and
A. Gal-Yam\inst{2} \href{https://orcid.org/0000-0002-3653-5598}{\includegraphics[scale=0.05]{figures/ORCIDiD_iconvector.eps}}\and
A. Gkini\inst{1} \href{https://orcid.org/0009-0000-9383-2305}{\includegraphics[scale=0.05]{figures/ORCIDiD_iconvector.eps}}\and
K. R. Hinds\inst{7} \href{https://orcid.org/0000-0002-0129-806X}{\includegraphics[scale=0.05]{figures/ORCIDiD_iconvector.eps}}\and
R. Lunnan\inst{1} \href{https://orcid.org/0000-0001-9454-4639}{\includegraphics[scale=0.05]{figures/ORCIDiD_iconvector.eps}}\and
D. Perley\inst{7} \href{https://orcid.org/000-0001-8472-1996}{\includegraphics[scale=0.05]{figures/ORCIDiD_iconvector.eps}}\and
YJ. Qin\inst{5} \href{https://orcid.org/0000-0003-3658-6026}{\includegraphics[scale=0.05]{figures/ORCIDiD_iconvector.eps}}\and
R. Stein\inst{5} \href{https://orcid.org/0000-0003-2434-0387}{\includegraphics[scale=0.05]{figures/ORCIDiD_iconvector.eps}}\and
J. Wise\inst{7} \href{https://orcid.org/0000-0003-0733-2916}{\includegraphics[scale=0.05]{figures/ORCIDiD_iconvector.eps}}\and
L. Yan\inst{8} \href{https://orcid.org/0000-0003-1710-9339}{\includegraphics[scale=0.05]{figures/ORCIDiD_iconvector.eps}}\and
E. A. Zimmerman\inst{2} \href{https://orcid.org/0000-0001-8985-2493}{\includegraphics[scale=0.05]{figures/ORCIDiD_iconvector.eps}}\and
S. Anand\inst{5} \href{https://orcid.org/0000-0003-3768-7515}{\includegraphics[scale=0.05]{figures/ORCIDiD_iconvector.eps}}\and
R. J. Bruch\inst{2} \href{https://orcid.org/0000-0001-8208-2473}{\includegraphics[scale=0.05]{figures/ORCIDiD_iconvector.eps}}\and
R. Dekany\inst{8} \href{https://orcid.org/0000-0002-5884-7867}{\includegraphics[scale=0.05]{figures/ORCIDiD_iconvector.eps}}\and
A. J. Drake\inst{5} \href{https://orcid.org}{\includegraphics[scale=0.05]{figures/ORCIDiD_iconvector.eps}}\and
C. Fremling\inst{8} \href{https://orcid.org/0000-0002-4223-103X}{\includegraphics[scale=0.05]{figures/ORCIDiD_iconvector.eps}}\and
B. Healy\inst{9} \href{https://orcid.org/0000-0002-7718-7884}{\includegraphics[scale=0.05]{figures/ORCIDiD_iconvector.eps}}\and
V. Karambelkar\inst{5} \href{https://orcid.org/0000-0003-2758-159X}{\includegraphics[scale=0.05]{figures/ORCIDiD_iconvector.eps}}\and
M. M. Kasliwal\inst{5} \href{https://orcid.org/0000-0002-5619-4938}{\includegraphics[scale=0.05]{figures/ORCIDiD_iconvector.eps}}\and
M. Kong\inst{5} \href{https://orcid.org}{\includegraphics[scale=0.05]{figures/ORCIDiD_iconvector.eps}}\and
  S. R. Kulkarni\inst{5} \href{https://orcid.org/0000-0001-5390-8563}{\includegraphics[scale=0.05]{figures/ORCIDiD_iconvector.eps}}\and
F. J. Masci\inst{10} \href{https://orcid.org/0000-0002-8532-9395}{\includegraphics[scale=0.05]{figures/ORCIDiD_iconvector.eps}}\and
R. S. Post\inst{11} \href{https://orcid.org/0000-0003-3244-0337}{\includegraphics[scale=0.05]{figures/ORCIDiD_iconvector.eps}}\and
J. Purdum\inst{8} \href{https://orcid.org/0000-0003-1227-3738}{\includegraphics[scale=0.05]{figures/ORCIDiD_iconvector.eps}}\and
R. Michael Rich\inst{12} \href{https://orcid.org/0000-0003-0427-8387}{\includegraphics[scale=0.05]{figures/ORCIDiD_iconvector.eps}}\and
A. Wold\inst{10} \href{https://orcid.org/0000-0002-9998-6732}{\includegraphics[scale=0.05]{figures/ORCIDiD_iconvector.eps}}} %
\institute{
The Oskar Klein Centre, Department of Astronomy, Stockholm University, AlbaNova, SE-106 91 Stockholm, Sweden 
  \and 
 Department of Particle Physics and Astrophysics, Weizmann Institute of Science, 234 Herzl St, 7610001 Rehovot, Israel 
  \and 
 Center for Interdisciplinary Exploration and Research in Astrophysics (CIERA), Northwestern University, 1800 Sherman Ave., Evanston, IL 60201, USA 
  \and 
 The Oskar Klein Centre, Department of Physics, Stockholm University, Albanova University Center, 106 91 Stockholm, Sweden 
  \and 
 Division of Physics, Mathematics and Astronomy, California Institute of Technology, Pasadena, CA 91125, USA 
  \and 
 MIT-Kavli Institute for Astrophysics and Space Research, 77 Massachusetts Ave., Cambridge, MA 02139, USA 
  \and 
 Astrophysics Research Institute, Liverpool John Moores University, IC2, Liverpool Science Park, 146 Brownlow Hill, Liverpool L3 5RF, UK 
  \and 
 Caltech Optical Observatories, California Institute of Technology, Pasadena, CA 91125, USA 
  \and 
 School of Physics and Astronomy, University of Minnesota, Minneapolis, MN 55455, USA 
  \and 
 IPAC, California Institute of Technology, 1200 E. California Blvd, Pasadena, CA 91125, USA 
  \and 
 Post Observatory, Lexington, MA 02421, USA 
  \and 
 Department of Physics \& Astronomy, University of California Los Angeles, PAB 430 Portola Plaza, Los Angeles, CA 90095-1547, USA
 } 
\date{}

\abstract
{Obtaining spectroscopic observations of the progenitors of core-collapse supernovae is often unfeasible, due to an inherent lack of knowledge as to what stars will go supernova and when they will explode. In this Letter, we present photometric and spectroscopic observations of the progenitor activity of \fyq before the He-rich progenitor explodes as a Type Ibn supernova. The progenitor of \fyq shows an exponential rise in flux prior to core-collapse. Complex \ion{He}{I} emission line features are observed in the progenitor spectra, with a P-Cygni like profile, as well as an evolving broad base with velocities on the order of 10,000~\kms. The luminosity and evolution of \fyq is consistent with a Type Ibn, reaching a peak $r$-band magnitude of $-18.8$~mag, although there is some uncertainty in the distance to the host, NGC 4388, located in the Virgo cluster. We present additional evidence of asymmetric He-rich material being present prior to the explosion of \fyq, as well as after, suggesting this material has survived the ejecta-CSM interaction. Broad [\ion{O}{I}], \ion{C}{I}, and the \ion{Ca}{II} triplet lines are observed at late phases, confirming that \fyq was a genuine supernova, rather than a non-terminal interacting transient. \fyq provides insight into the final moments of a massive star's life, highlighting that the progenitor is likely highly unstable before core-collapse. 
}

         
\maketitle

\section{Introduction}\label{sec:introduction}

Massive stars ($\gtrsim$ 8 -- 10~\msun) will eventually end their lives as core-collapse supernovae \citep[CCSNe; ][]{Woosley1995,Heger2003,Janka2012,Crowther2012}. With the increasing capabilities of photometric transient surveys \citep{Bellm2014,Chambers2016,Tonry2018} a growing sample of supernova (SN) progenitors have been observed experiencing outbursts in the weeks -- years prior to their ultimate core-collapse \citep{Foley2007, Pastorello2007, Mattila2008,Ofek2013,Fraser2013,Margutti2014,Ofe2014,Strotjohann2021,Fransson2022,Jacobson2022, Smith2022,Hiramatsu2023}. These precursor events are difficult to explain, and are a relatively new area of stellar astrophysics \citep{Dessart2010,Quataert2012,Leung2021,Strotjohann2021,Tsang2022}, however are often invoked to explain complex circumstellar material (CSM) shortly before collapse \citep{Matsumoto2022,Hiramatsu2023}. Precursor activity is likely accompanied by violent mass ejections by a currently unclear mechanism. Potential scenarios include binary interaction \citep{Yoon2010,Smith2011,Kashi2013,Yoon2017,Zapartas2021,Tsuna2024}, outbursts from Luminous Blue Variables \citep[LBVs; ][]{Humphreys1994,Smith2010, Kilpatrick2018,Smith2022}, and instabilities towards the end of a massive stars life \citep{Woosley2007,Smith2014a,Muller2016,Fuller2018,Leung2021,Wu2022}.

\begin{figure}
    \centering
    \includegraphics[width= \columnwidth]{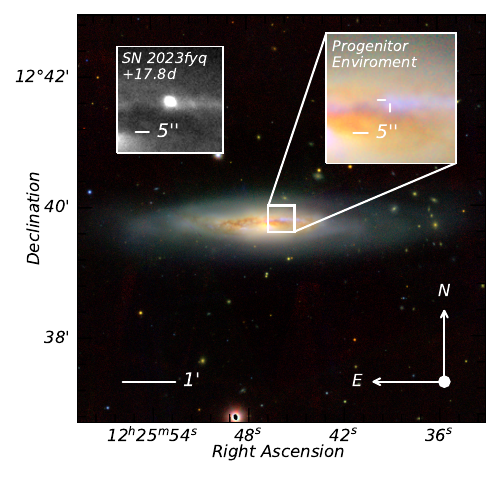}
    \caption{Composite $giy$ color image of the host galaxy of \fyq from the Pan-STARRS survey, with the transient environment given in the top right inset. An $r$-band image of \fyq from the NOT/ALFOSC from 2023-08-16 is given in the top left inset.
    }
    \label{fig:finder}
\end{figure}

This Letter presents spectra (photometric) observations of the progenitor of a Type Ibn SN several months (years) before core-collapse, as well as \fyq itself. The precursor activity of \fyq\footnote{Also known as ZTF22abzzvln, ATLAS23rwh, and PS23fnw.} was discovered by the Zwicky Transient Facility \citep[ZTF; ][]{Graham2019,Bellm2019,Masci2019,Dekany2020} at  R.A. = 12:25:45.874, Dec. = $+$12:39:48.87 (J2000) on 2023-04-17 at $g=19.51$~mag (AB) \citep{De2023}. Spectroscopic observations of the progenitor were obtained as part of the  Census of the Local Universe (CLU) survey \citep{De2020} as it satisfied the selection criteria; low-luminosity transient in a nearby galaxy. \fyq was later classified by the Nordic optical telescope Unbiased Transient Survey 2 (NUTS2) as a Type Ib-pec SN \citep{Valerin2023} on 2023-07-25, around 5 days pre-peak. \fyq occurred around 10$''$ from the core of  the active spiral galaxy NGC 4388 \citep{Damas-Segovia2016} in the Virgo Cluster, as presented in Fig. \ref{fig:finder}. For consistency with the literature, we assume a distance of 17.2~Mpc to NGC 4388 \citep{Lianou2019,Bohringer1997}, and a distance modulus ($\mu$) of 31.2~mag, giving \fyq a peak $r$-band magnitude of $-18.8$~mag. However the distance is uncertain and may be as high as 25~Mpc \citep[e.g. ][]{Ekholm2000}, meaning \fyq may be $\sim1.5$~mag more luminous. We correct for foreground Milky Way (MW) extinction using ${\rm R_V}=3.1$, ${\rm E (B - V)_{MW}}$ = 0.029~mag \citep{Schlafly2011}, with the extinction law given by \citet{Cardelli1989}. Measurements of host emission lines indicates substantial host attenuation, as inferred from the Balmer (\halpha/\hbeta) decrement, of ${\rm E(B-V)_{host}\approx0.4\pm0.1}$~mag \citep{Osterbrock1989}. We correct for this extinction in a similar fashion to Milky Way extinction mentioned above using an ${\rm R_V}=2$. We adopt a redshift of $z = 0.0084$ and the rest-frame phase is reported with respect to the $r$-band peak magnitude on 2023-07-30 ({$t_{peak}$} = 60155.1).

\section{Observations}\label{sec:obs}

\subsection{Photometric observations}\label{ssec:obsdata_photometry}

Science-ready images taken as part of the ZTF survey were obtained in $gri$ from the NASA/IPAC Infrared Science Archive (IPAC\footnote{\href{https://irsa.ipac.caltech.edu/applications/ztf/}{https://irsa.ipac.caltech.edu/applications/ztf/}}) service. Template subtracted photometry was performed using the \software{AutoPhOT} pipeline \citep{Brennan2022a}. Photometry from the ATLAS forced-photometry server\footnote{\href{https://fallingstar-data.com/forcedphot/}{https://fallingstar-data.com/forcedphot/}} \citep{Tonry2018,Smith2020,Shingles2021a} were obtained in $c$ and $o$ bands. Several epochs of optical photometry were obtained with the Alhambra Faint Object Spectrograph and Camera (ALFOSC) on the 2.56~m Nordic Optical Telescope (NOT) in $gri$, the Liverpool Telescope (LT) with the optical imaging component of Infrared-Optical imager: Optical (IO:O) in $griz$, and the Spectral Energy Distribution Machine \citep[SEDM;][]{Blagorodnova2018,Rigault2019} on the Palomar 60-inch (P60) telescope in $gri$. We also obtained $gri$ images from a 32-inch Ritchey-Chretien telescope (RC32) at Post Observatory in Mayhill, New Mexico. UV photometry were acquired using the Ultra-violet Optical Telescope (UVOT) onboard the \textit{Neil Gehrels Swift Observatory} \citep{Gehrels2004,Roming2005}. We reduced the images using the \textit{Swift} \software{HEAsoft}\footnote{\href{https://heasarc.gsfc.nasa.gov/docs/software/heasoft/}{https://heasarc.gsfc.nasa.gov/docs/software/heasoft/}}  V. 6.26.1. toolset, as detailed in \citet{Irani2023}. A UV template image was constructed using archival observations prior to MJD 59100 and after the SN declined to the host level. We then removed the local host-galaxy contribution by subtracting the SN site flux from the fluxes of the individual epochs.

\subsection{Spectroscopic observations}\label{ssec:obsdata_spectroscopy}

Follow-up spectra were obtained using NOT/ALFOSC, the Low Resolution Imaging Spectrometer \citep[LRIS; ][]{Oke1995} on the 10m Keck telescope, the Double Spectrograph (DBSP) on the Palomar 200-inch telescope (P200), the Kast spectrograph at the Lick Observatory, the Gemini Multi-Object Spectrographs (GMOS) mounted on the Gemini North 8m telescope \citep{Hook2004}, and P60/SEDM \citep{Blagorodnova2018}. Spectra were reduced in a standard manner, using \software{LPipe}  \citep{Perley2019}, \software{DBSP\_DRP} \citep{Mandigo2022} and \software{PypeIt} \citep{pypeit:zenodo,pypeit:joss_arXiv,pypeit:joss_pub}, for Keck/LRIS, P200/DBSP, and NOT/ALFOSC, \software{pysedm} \citep{Rigault2019,Kim2022} for SEDM spectra, the \software{UCSC} spectral pipeline\citep{Siebert2019} for the Kast/Lick spectrum, and the \software{DRAGONS} \citep{Labrie2019} for our GEMINI/GMOS spectrum. Observations using the P60, P200,and Keck telescopes were coordinated using the FRITZ data platform \citep{Walt2019,Coughlin2023}. 

NGC 4388 was observed with the Multi Unit Spectroscopic Explorer (MUSE) at the 8.2~m ESO Very Large Telescope on 12 March 2022 (ID. 108.229J, PI. Venturi). Observations were reduced using the MUSE pipeline v2.8.7 and the ESO workflow engine ESOReflex \citep{Freudling2013a, Weilbacher2020a}. We extracted all spectra with the MUSE Python Data Analysis Framework (MPDAF) version 3.6 \citep{Bacon2016a}. The spectrum of the SN site was extracted with a circular aperture. The aperture radius of $0.5''$  translates to a physical scale of 89.5~pc at the projected distance of \fyq, comparable to the largest giant molecular clouds in the Milky Way.

\begin{figure*}
    \centering
    \includegraphics[width= \textwidth]{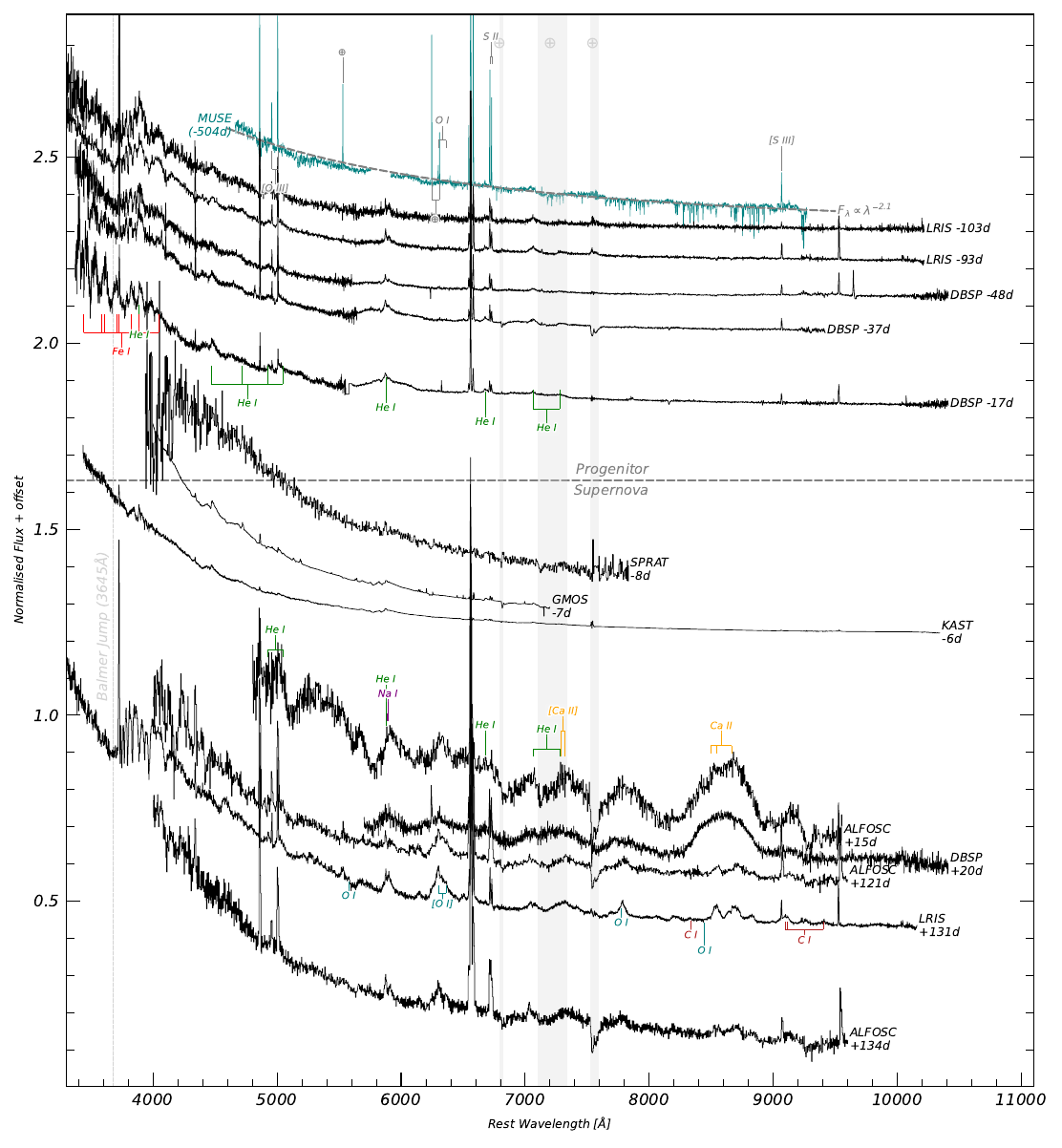}
    \caption{Spectral observations of \fyq and its progenitor. Each spectrum has been normalised and offset for clarity, as well as corrected for redshift and extinction. We mark regions of telluric contamination using the $\oplus$ symbol and grey vertical bands.}
    \label{fig:spectra}
\end{figure*}

Spectroscopic observations for the progenitor activity and \fyq are presented in Fig. \ref{fig:spectra}.

\section{Results and Discussion}

\subsection{The photometric evolution of \fyq and its progenitor activity}

\begin{figure*}[!ht]
    \centering
    \includegraphics[width= \textwidth]{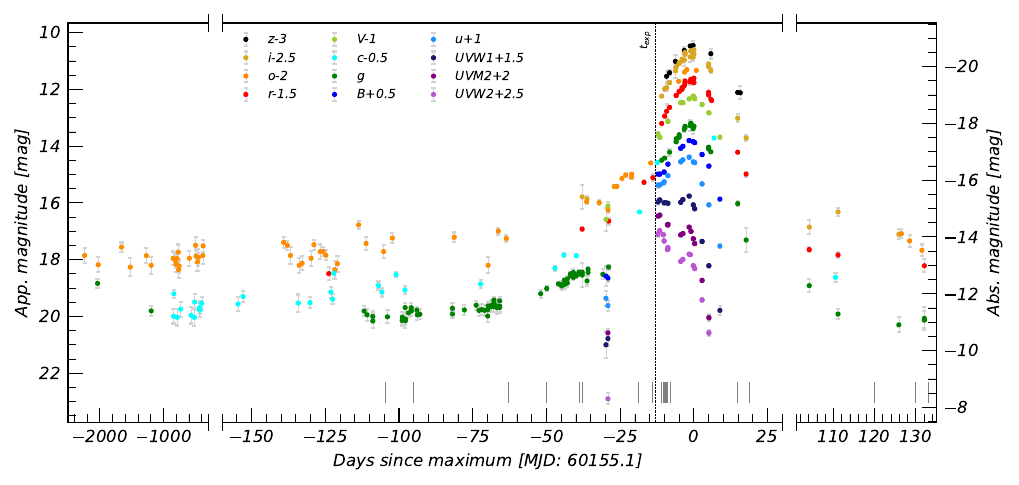}
    \caption{Multiband light curve of \fyq, with each band offset for clarity.  Vertical grey bars denote epochs of spectral observations. Absolute magnitudes, based on a distance of 17.2 Mpc, are shown on the right Y-axis, uncorrected for Milky Way or host extinction. Note the broken X-axis, highlighting the long-lived progenitor activity.}
    \label{fig:lightcurve}
\end{figure*}

\begin{figure}[!ht]
    \centering
    \includegraphics[width= \columnwidth]{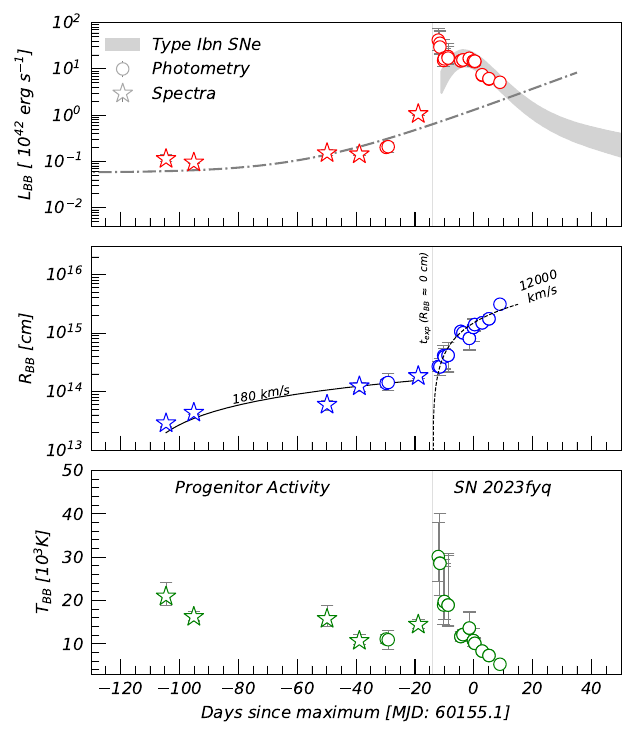}
    \caption{The blackbody luminosity of \fyq and its progenitor are depicted in the figure. Star markers indicate blackbody fits to the synthetic photometry from spectra presented in Fig. \ref{fig:spectra}. Circle markers are blackbody fits to epochs of photometry which contain at least $u$-band. Similarly, blackbody radius and temperature are provided in the middle and lower panels, respectively.}
    \label{fig:bolometric_lightcurve}
\end{figure}

Figure \ref{fig:lightcurve} gives the photometric evolution of \fyq, highlighting the pre-SN evolution which lasts around 5 years, as well as the SN explosion itself. Figure \ref{fig:bolometric_lightcurve} provides the bolometric evolution constructed from the host-subtracted photometry presented in Fig.~\ref{fig:lightcurve}, and calibrated using the extinction and distance given in Sect.~\ref{sec:introduction}. 

The blackbody luminosity (${\rm L_{BB}}$) of the progenitor is well-fitted by an accelerating rise (i.e., exponential), followed by a fast rise after the progenitor has likely exploded. The blackbody radius (${\rm R_{BB}}$) of the progenitor is seen to increase linearly at approximately 200~\kms, peaking at ${\rm 1.75\times10^{14}~cm}$, after which a sharp increase is observed. This radius ($\sim$ 2400 ${\rm R_{\odot}}$) is unlikely to represent a photosphere of the progenitor star, but perhaps signifies an extended/inflated envelope, due to the final stages of a binary merger event \citep{Ivanova2013,Irani2023}, or a confined CSM, as this value is consistent with the shock breakout radius reported for SN 2023ixf \citep{Zimmerman2023}. Assuming this second increase is a result of core-collapse, we extrapolated backwards to ${\rm R_{BB} \approx 0 cm}$ and find an explosion time (${t_{\text{exp}}}$) of -13.9 days. Due to our sparse photometry around the explosion epoch, we are uncertain whether \fyq follows the evolution depicted by the analytical fits in Fig. \ref{fig:bolometric_lightcurve}, or rather a multi-component rise \citep[e.g., SN 2021qqp, \ip;][]{Hiramatsu2023}. We adopt a rise time for \fyq of approximately 14 days, which is consistent with what is seen in other Type Ibn SNe \citep{Hosseinzadeh2017}.

 From \fyq, we measure a max luminosity of ${\rm4.2\times 10^{43}~erg~s^{-1}}$, ${\rm L(t = t_{peak}) \approx 2.2\times 10^{41}~erg~s^{-1}}$, and a total radiated energy of ${\rm 6.0\times 10^{49}~erg}$. These value is consistent with predicted activity from binary interactions \citep{Tsuna2024}, and similar to progenitor activity in other SNe \citep{Strotjohann2021}. We measure the radiated energy of the pre-SN evolution as ${\rm L(t < t_{exp}) \approx 6.6\times 10^{47}~erg}$. Due to the lack of UV and IR coverage of the progenitor and \fyq, these values are taken as lower limits. After the SN has occurred, a dramatic increase in ${\rm L_{BB}}$ and ${\rm T_{BB}}$ is observed, with temperatures reaching $\sim 30~\text{kK}$. This brightness increase is mainly seen in the bluer bands (see Fig. \ref{fig:lightcurve}), and is consistent with shock breakout \citep[e.g., SN 2008D;][]{Chevalier2008, Modjaz2009}.

Assuming an efficient conversion of kinetic energy to radiation, we get ${ \rm M_{CSM} \approx\frac{2E_{rad}}{\epsilon v^2} \approx 0.05 }$~\msun, assuming 50\% energy conversion, and a velocity of 1700~\kms (taken from the observed P-Cygni minimum in pre-SN spectra). This material would be very optically thick at $10^{14}$~cm, which is consistent with the observed thermalised spectra. For \fyq, a CSM breakout with $\sim$0.05~\msun would last a few days, with a peak luminosity of a few $10^{43}$ erg~s$^{-1}$ followed by shock emergence and cooling, consistent with Fig. \ref{fig:bolometric_lightcurve}. In this scenario \citep[][e.g. Edge-breakout; Heavy CSM scenerio ]{Khatami2023}, we expect ${\rm M_{ejecta} > M_{CSM}}$, and given the higher mass and lower velocity of the shocked CSM material, the cooling emission diffuses out on a longer timescale \citep{Piro2015}. We estimate the diffusion time \citep[taken to be ${ \rm t_d \approx \sqrt{\frac{\kappa M_{CSM}}{vc}} }$; ][]{Khatami2019} as $\sim$12 days, assuming material expanding at 12000~\kms from Fig. \ref{fig:bolometric_lightcurve}, roughly consistent with the adopted explosion time. Assuming our final $R_{BB}$ measurement before the adopted ${ \rm t_{exp} }$ indicates the radius of this CSM material, we find ${\rm Log(R/R_{\odot})} \approx 3.4$, which overall agrees with models from \citet{{Khatami2023}}.

We measure the progenitor to have ${\rm \log(T_{eff} / K)} \approx 4.1 \pm 0.1$ and ${\rm \log(L / L_{\odot})} \approx 7.6 \pm 0.4$ from host-subtracted photometry. These values are likely super-Eddington and are reminiscent of the Giant Eruption of Eta Carina, which, possibly through multi-star interactions, ejected $\sim$10 \msun of material but did not unbind the star \citep{Smith2011, Prieto2014}. This suggests a multi star scenario, although without detailed late-time stellar evolutionary modeling, the mechanism powering the pre-SN evolution is unclear.

\subsection{Spectroscopic evolution}\label{ssec:spectro_evo}

\begin{figure}[!ht]
    \centering
    \includegraphics[width= \columnwidth]{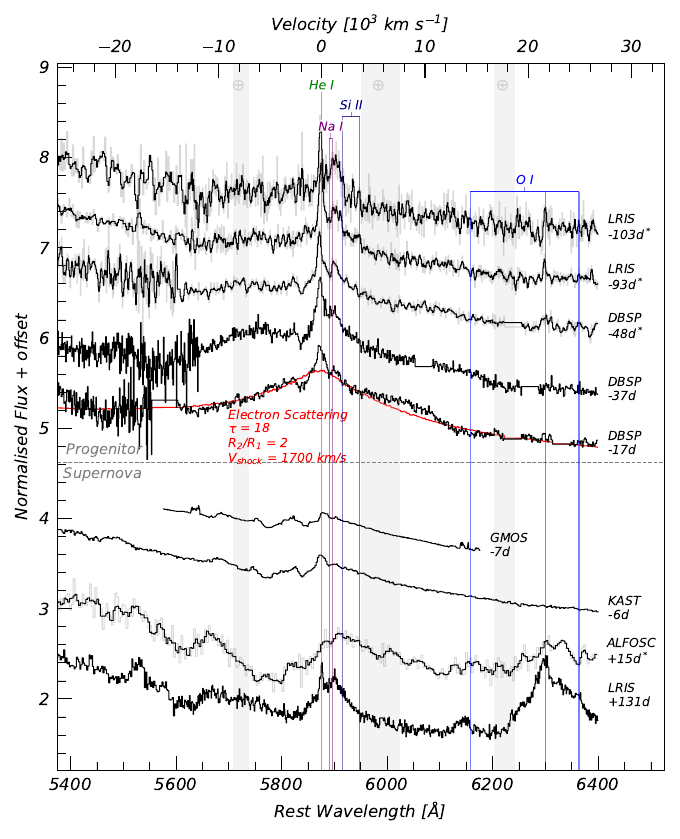}
    \caption{Evolution of the \ion{He}{I} $\lambda$5876 emission profile. Pre-SN spectra show complex P-Cygni like profiles. Soon after the SN has occurred, an additional absorption component is seen in the $-$6d Lick/KAST spectrum at around $-$5000~\kms. At late times, the P-Cygni absorption is no longer observed, but rather a double-peaked emission profile consisting of host emission and redshifted \ion{He}{I} emission at similar velocities as seen in the $-$93d Keck/LRIS spectrum. Spectra marked with an asterisk have been re-binned for clarity, with the raw spectrum given in grey. We manually mask out telluric lines and artefacts from the sky subtractions.}
    \label{fig:helium_evolution}
\end{figure}

Observations of CCSN progenitors are rare, with the only previous examples, to our knowledge, being  \ip \citep{Smith2010,Foley2011,Pastorello2013} and \bh \citep{Elias-Rosa2016,Thone2017}. In agreement with stellar evolutionary theory, the progenitor of the Type Ibn \fyq display features expected for He-rich stars \citep{Crowther2007,Yoon2010}. Figure \ref{fig:helium_evolution} shows the evolution of \ion{He}{I} $\lambda$5876, during the pre- and post- SN epochs, as well as the post-peak appearance. Similar to the \halpha profile seen for \ip\footnote{Although \ip is a H-rich Type IIn SN and \fyq a H-poor Type Ibn SN, we make qualitative comparisons between the appearance of their progenitor star/environments.} in 2009 \citep{Pastorello2013}, \fyq shows a complex P-Cygni like profile for \ion{He}{I} $\lambda$5876 (and similarly for other \ion{He}{I} lines) with an absorption centered at $-1700$~\kms and extending out to $-2500$~\kms. Perhaps similarly to \halpha in late time spectra of \ip \citep{Fraser2015}, the \ion{He}{I} in \fyq shows a redshifted emission peak, centered at +1400~\kms,  which is present during the slow pre-SN rise, as well as post explosion. The velocities of these profiles do no evolve significantly during the pre-SN stage, although this may be a result of poor S/N.

A spectrum of the explosion site from VLT/MUSE taken $\sim$2 years prior to \fyq is given in Fig. \ref{fig:spectra}. We investigate if any potential flux from the progenitor activity is present in this spectrum by comparing it to a similar spectrum extracted 2.4$''$. We note a tentative broad excess at the position of \ion{He}{I} $\lambda\lambda$ 6678,7065, possibly due to progenitor emission. 

Due to the sodium laser guide star, the wavelength range covering \ion{He}{I} $\lambda$5876 is masked, preventing any investigation of potential host contamination of \ion{He}{I} $\lambda$5876 for \fyq. We suspect this redshifted component is originating from \ion{He}{I} $\lambda$5876 as it is seen in several \ion{He}{I} lines in the pre-SN spectra, as well as a similar emission component at similar velocities seen in the post-SN spectra. This would mean that the asymmetric material responsible for this emission was not destroyed in the SN explosion. SN ejecta interacting with asymmetric CSM has been used to explain irregular emission line profiles \citep[e.g. ][]{Leloudas2015,Andrews2018,Pursiainen2022,Hosseinzadeh2022,Smith2023a} and bumpy light curves \citep{Nyholm2017,Woosley2018,Wang2022}, and \fyq provides the first clear spectroscopic evidence of asymmetric structure prior to core-collapse.

A broad component is observed for \ion{He}{I} $\lambda$5876 profile in the $-37$d and $-17$d spectra, with wings extending to $\pm13,000$~\kms and a full width half maximum (FWHM) around 14,000~\kms. Such velocities are normally associated with SN ejecta, although have also been observed in non-terminal eruptions \citep{Pastorello2013,Smith2018}. To investigate if this broad component evolves, we measure the flux in a small wavelength bin around 5800~\AA\ in the continuum-subtracted flux-calibrated spectra. An order of magnitude increase in flux from $-103$ to $-10$ days is observed. This confirms that the broad component is rapidly evolving in strength prior to the explosion of \fyq rather than not being detectable in earlier spectra due to low signal-to-noise. 

While these features may represent fast moving material, an alternative explanation could be electron scattering. Some mechanism(s) cause the progenitor to be surrounded by a dense CSM \citep[such as acoustic waves or a merger;][]{Smith2011, Yoon2017,Fuller2018,Chevalier2012,Fransson2022,Tsuna2024}, and may lead to shock dissipation and emission of radiation in the optically thick CSM. We model the $-17$~d spectrum using the electron scattering model from \citet{Brennan2023}, assuming ${\rm V_{\text{shock}} = 1700}$~\kms, ${\rho \propto r^{-2}}$, ${\rm \tau_{e} \approx 18}$, ${ \rm T_e = 2\times 10^{4}}$~K, and ${ \rm R_2 = 1.75 \times 10^{14}}$~cm. As shown in Fig. \ref{fig:helium_evolution}, the model reproduces the wings of \ion{He}{I} $\lambda$5876 well, suggesting that under certain conditions, material moving at $>10^4$~\kms may not be required to produce the broad features during the pre-SN phase. Assuming a random walk, the diffusion time is $t_d = \frac{R^2}{\lambda_{\text{mfp}}c} \approx 4$~days. Our models do not produce the P-Cygni-like profile or offset emission and are poorly fit in the red wing, highlighting the complex geometry likely responsible for these features.

The post-SN \ion{He}{I} $\lambda$5876 appears double peaked with similar shaped profiles also seen in the ${\ion{Ca}{II}~\lambda\lambda8498, 8542, 8662}$ near-infrared triplet. Double peaked emission lines have been observed in Type Ibn SNe \citep[e.g. SN~2018bcc][]{Karamehmetoglu2021}, as well as Type IIn SNe \citep{Brennan2022b,Hiramatsu2023}, and are attributed to asymmetric SN ejecta or CSM or both \citep{Andrews2019, Pursiainen2022}.

A long standing question for \ip-like transients is whether the transient is indeed a SN or rather a ``SN impostor'' \citep{VanDyk2000,Mauerhan2013,Fraser2013,Brennan2022c,Smith2022}. As shown in Figs.~\ref{fig:spectra} and \ref{fig:helium_evolution}, \fyq shows broad emission from ${\rm [\ion{O}{I}]~\lambda\lambda6300, 6364}$ at nebular phases, showing that \fyq is a genuine CCSN \citep{Jerkstrand2014,Kuncarayakti2015}, whose progenitor underwent some form of instabilities shortly before core-collapse. Late time followup of \fyq make reveal significant reddening due to possible dust formation. We note the appearance of \ion{O}{I} and \ion{C}{I} emission lines, meaning the environment of \fyq is conducive for amorphous and graphite dust grains.

Events like \fyq and \ip suggest that certain massive stars experience eruptive activity preceding core-collapse. However, several nearby transients lack notable eruptive activity \citep{Jacobson2022,Ransome2023,Dong2023}, implying a diverse mechanisms for pre-supernova emissions in the final stages before core-collapse.

\section{Conclusion}

In this Letter, we present an unprecedented dataset of progenitor activity of a Type Ibn SN almost 150 days before core-collapse occurs. Pre-SN spectra reveal a complex evolving \ion{He}{I} profile. These observations of \fyq and the final moments of the progenitor, highlight that the progenitors to CCSNe can undergo some extreme instabilities shortly before their final demise. 

Progenitor analysis typically occurs after the star has been destroyed, by searching through archival images, and measuring the photometric properties of the assumed progenitor. Although this area of transient astronomy is in its infancy, the repercussions of detecting precursor activity are immense, highlighting that the progenitor is not in a equilibrium state, and may not be represented well by standard stellar evolutionary models. \fyq and similar transients, highlight that the pre-SN appearance of the progenitor is non-trivial, and without careful consideration, may produce misleading results in supernova progenitor studies.

\section*{Data availability}

The spectroscopic and photometric data underlying this article are available in the Weizmann Interactive Supernova Data Repository  \citep[WISeREP\footnote{\href{https://wiserep.weizmann.ac.il/}{https://wiserep.weizmann.ac.il/}};][]{Yaron2012}.

\section*{Acknowledgements}

We thank the anonymous referee for their comments on host extinction and CSM mass, which have improved the paper. SJB would like to thank M. Fraser and A. Guinness for their insights into late time stellar evolution and precursor events. 
SJB and RL acknowledges their support by the European Research Council (ERC) under the European Union’s Horizon Europe research and innovation programme (grant agreement No. 10104229 - TransPIre).
 S. Schulze is partially supported by LBNL Subcontract NO. 7707915 and by the G.R.E.A.T. research environment, funded by Vetenskapsr{\aa}det,  the Swedish Research Council, project number 2016-06012.  
Based on observations obtained with the Samuel Oschin Telescope 48-inch and the 60-inch Telescope at the Palomar Observatory as part of the Zwicky Transient Facility project. ZTF is supported by the National Science Foundation under Grant No. AST-2034437 and a collaboration including Caltech, IPAC, the Weizmann Institute of Science, the Oskar Klein Center at Stockholm University, the University of Maryland, Deutsches Elektronen-Synchrotron and Humboldt University, the TANGO Consortium of Taiwan, the University of Wisconsin at Milwaukee, Trinity College Dublin, Lawrence Livermore National Laboratories, IN2P3, University of Warwick, Ruhr University Bochum and Northwestern University. Operations are conducted by COO, IPAC, and UW. SED Machine is based upon work supported by the National Science Foundation under Grant No. 1106171. This work was supported by the GROWTH project funded by the National Science Foundation under Grant No 1545949. The Oskar Klein Centre is funded by the Swedish Research Council. The data presented here were obtained in part with ALFOSC, which is provided by the instituto de Astrofisica de Andalucia (IAA) under a joint agreement with the University of Copenhagen and NOT. The ZTF forced-photometry service was funded under the Heising-Simons Foundation grant \#12540303 (PI: Graham). The Gordon and Betty Moore Foundation, through both the Data-Driven Investigator Program and a dedicated grant, provided critical funding for SkyPortal. The Liverpool Telescope is operated on the island of La Palma by Liverpool John Moores University in the Spanish Observatorio del Roque de los Muchachos of the Instituto de Astrofisica de Canarias with financial support from the UK Science and Technology Facilities Council. Based on observations obtained at the international Gemini Observatory, a program of NSF’s NOIRLab, which is managed by the Association of Universities for Research in Astronomy (AURA) under a cooperative agreement with the National Science Foundation on behalf of the Gemini Observatory partnership: the National Science Foundation (United States), National Research Council (Canada), Agencia Nacional de Investigaci\'{o}n y Desarrollo (Chile), Ministerio de Ciencia, Tecnolog\'{i}a e Innovaci\'{o}n (Argentina), Minist\'{e}rio da Ci\^{e}ncia, Tecnologia, Inova\c{c}\~{o}es e Comunica\c{c}\~{o}es (Brazil), and Korea Astronomy and Space Science Institute (Republic of Korea).  The Gemini observations were obtained through program GN-2023A-Q-122 and processed using DRAGONS (Data Reduction for Astronomy from Gemini Observatory North and South). This work was enabled by observations made from the Gemini North telescope, located within the Maunakea Science Reserve and adjacent to the summit of Maunakea. We are grateful for the privilege of observing the Universe from a place that is unique in both its astronomical quality and its cultural significance.

\bibliographystyle{aa}
\bibliography{references.bib} %

\begin{thebibliography}{112}
\expandafter\ifx\csname natexlab\endcsname\relax\def\natexlab#1{#1}\fi

\bibitem[{{Andrews} {et~al.}(2019){Andrews}, {Sand}, {Valenti}, {Smith},
  {Dastidar}, {Sahu}, {Misra}, {Singh}, {Hiramatsu}, {Brown}, {Hosseinzadeh},
  {Wyatt}, {Vinko}, {Anupama}, {Arcavi}, {Ashall}, {Benetti}, {Berton},
  {Bostroem}, {Bulla}, {Burke}, {Chen}, {Chomiuk}, {Cikota}, {Congiu}, {Cseh},
  {Davis}, {Elias-Rosa}, {Faran}, {Fraser}, {Galbany}, {Gall}, {Gal-Yam},
  {Gangopadhyay}, {Gromadzki}, {Haislip}, {Howell}, {Hsiao}, {Inserra},
  {Kankare}, {Kuncarayakti}, {Kouprianov}, {Kumar}, {Li}, {Lin}, {Maguire},
  {Mazzali}, {McCully}, {Milne}, {Mo}, {Morrell}, {Nicholl}, {Ochner},
  {Olivares}, {Pastorello}, {Patat}, {Phillips}, {Pignata}, {Prentice},
  {Reguitti}, {Reichart}, {Rodr{\'\i}guez}, {Rui}, {Sanwal}, {S{\'a}rneczky},
  {Shahbandeh}, {Singh}, {Smartt}, {Strader}, {Stritzinger}, {Szak{\'a}ts},
  {Tartaglia}, {Wang}, {Wang}, {Wang}, {Wheeler}, {Xiang}, {Yaron}, {Young}, \&
  {Zhang}}]{Andrews2019}
{Andrews}, J.~E., {Sand}, D.~J., {Valenti}, S., {et~al.} 2019, \apj, 885, 43

\bibitem[{{Andrews} \& {Smith}(2018)}]{Andrews2018}
{Andrews}, J.~E. \& {Smith}, N. 2018, \mnras, 477, 74

\bibitem[{{Bacon} {et~al.}(2016){Bacon}, {Piqueras}, {Conseil}, {Richard}, \&
  {Shepherd}}]{Bacon2016a}
{Bacon}, R., {Piqueras}, L., {Conseil}, S., {Richard}, J., \& {Shepherd}, M.
  2016, {MPDAF: MUSE Python Data Analysis Framework}, Astrophysics Source Code
  Library, record ascl:1611.003

\bibitem[{{Bellm}(2014)}]{Bellm2014}
{Bellm}, E. 2014, in The Third Hot-wiring the Transient Universe Workshop, ed.
  P.~R. {Wozniak}, M.~J. {Graham}, A.~A. {Mahabal}, \& R.~{Seaman}, 27--33

\bibitem[{{Bellm} {et~al.}(2019){Bellm}, {Kulkarni}, {Graham}, {Dekany},
  {Smith}, {Riddle}, {Masci}, {Helou}, {Prince}, {Adams}, {Barbarino},
  {Barlow}, {Bauer}, {Beck}, {Belicki}, {Biswas}, {Blagorodnova}, {Bodewits},
  {Bolin}, {Brinnel}, {Brooke}, {Bue}, {Bulla}, {Burruss}, {Cenko}, {Chang},
  {Connolly}, {Coughlin}, {Cromer}, {Cunningham}, {De}, {Delacroix}, {Desai},
  {Duev}, {Eadie}, {Farnham}, {Feeney}, {Feindt}, {Flynn}, {Franckowiak},
  {Frederick}, {Fremling}, {Gal-Yam}, {Gezari}, {Giomi}, {Goldstein},
  {Golkhou}, {Goobar}, {Groom}, {Hacopians}, {Hale}, {Henning}, {Ho}, {Hover},
  {Howell}, {Hung}, {Huppenkothen}, {Imel}, {Ip}, {Ivezi{\'c}}, {Jackson},
  {Jones}, {Juric}, {Kasliwal}, {Kaspi}, {Kaye}, {Kelley}, {Kowalski},
  {Kramer}, {Kupfer}, {Landry}, {Laher}, {Lee}, {Lin}, {Lin}, {Lunnan},
  {Giomi}, {Mahabal}, {Mao}, {Miller}, {Monkewitz}, {Murphy}, {Ngeow},
  {Nordin}, {Nugent}, {Ofek}, {Patterson}, {Penprase}, {Porter}, {Rauch},
  {Rebbapragada}, {Reiley}, {Rigault}, {Rodriguez}, {van Roestel}, {Rusholme},
  {van Santen}, {Schulze}, {Shupe}, {Singer}, {Soumagnac}, {Stein}, {Surace},
  {Sollerman}, {Szkody}, {Taddia}, {Terek}, {Van Sistine}, {van Velzen},
  {Vestrand}, {Walters}, {Ward}, {Ye}, {Yu}, {Yan}, \& {Zolkower}}]{Bellm2019}
{Bellm}, E.~C., {Kulkarni}, S.~R., {Graham}, M.~J., {et~al.} 2019, \pasp, 131,
  018002

\bibitem[{{Blagorodnova} {et~al.}(2018){Blagorodnova}, {Neill}, {Walters},
  {Kulkarni}, {Fremling}, {Ben-Ami}, {Dekany}, {Fucik}, {Konidaris}, {Nash},
  {Ngeow}, {Ofek}, {O' Sullivan}, {Quimby}, {Ritter}, \&
  {Vyhmeister}}]{Blagorodnova2018}
{Blagorodnova}, N., {Neill}, J.~D., {Walters}, R., {et~al.} 2018, \pasp, 130,
  035003

\bibitem[{{B{\"o}hringer} {et~al.}(1997){B{\"o}hringer}, {Neumann},
  {Schindler}, \& {Huchra}}]{Bohringer1997}
{B{\"o}hringer}, H., {Neumann}, D.~M., {Schindler}, S., \& {Huchra}, J.~P.
  1997, \apj, 485, 439

\bibitem[{{Brennan} \& {Fraser}(2022)}]{Brennan2022a}
{Brennan}, S.~J. \& {Fraser}, M. 2022, \aap, 667, A62

\bibitem[{{Brennan} {et~al.}(2022{\natexlab{a}}){Brennan}, {Fraser},
  {Johansson}, {Pastorello}, {Kotak}, {Stevance}, {Chen}, {Eldridge}, {Bose},
  {Brown}, {Callis}, {Cartier}, {Dennefeld}, {Dong}, {Duffy}, {Elias-Rosa},
  {Hosseinzadeh}, {Hsiao}, {Kuncarayakti}, {Martin-Carrillo}, {Monard},
  {Nyholm}, {Pignata}, {Sand}, {Shappee}, {Smartt}, {Tucker}, {Wyrzykowski},
  {Abbot}, {Benetti}, {Bento}, {Blondin}, {Chen}, {Delgado}, {Galbany},
  {Gromadzki}, {Guti{\'e}rrez}, {Hanlon}, {Harrison}, {Hiramatsu}, {Hodgkin},
  {Holoien}, {Howell}, {Inserra}, {Kankare}, {Koz{\l}owski},
  {M{\"u}ller-Bravo}, {Maguire}, {McCully}, {Meintjes}, {Morrell}, {Nicholl},
  {O'Neill}, {Pietrukowicz}, {Poleski}, {Prieto}, {Rau}, {Reichart},
  {Schweyer}, {Shahbandeh}, {Skowron}, {Sollerman}, {Soszy{\'n}ski},
  {Stritzinger}, {Szyma{\'n}ski}, {Tartaglia}, {Udalski}, {Ulaczyk}, {Young},
  {van Leeuwen}, \& {van Soelen}}]{Brennan2022b}
{Brennan}, S.~J., {Fraser}, M., {Johansson}, J., {et~al.} 2022{\natexlab{a}},
  \mnras, 513, 5642

\bibitem[{{Brennan} {et~al.}(2022{\natexlab{b}}){Brennan}, {Fraser},
  {Johansson}, {Pastorello}, {Kotak}, {Stevance}, {Chen}, {Eldridge}, {Bose},
  {Brown}, {Callis}, {Cartier}, {Dennefeld}, {Dong}, {Duffy}, {Elias-Rosa},
  {Hosseinzadeh}, {Hsiao}, {Kuncarayakti}, {Martin-Carrillo}, {Monard},
  {Pignata}, {Sand}, {Shappee}, {Smartt}, {Tucker}, {Wyrzykowski}, {Abbot},
  {Benetti}, {Bento}, {Blondin}, {Chen}, {Delgado}, {Galbany}, {Gromadzki},
  {Guti{\'e}rrez}, {Hanlon}, {Harrison}, {Hiramatsu}, {Hodgkin}, {Holoien},
  {Howell}, {Inserra}, {Kankare}, {Koz{\l}owski}, {M{\"u}ller-Bravo},
  {Maguire}, {McCully}, {Meintjes}, {Morrell}, {Nicholl}, {O'Neill},
  {Pietrukowicz}, {Poleski}, {Prieto}, {Rau}, {Reichart}, {Schweyer},
  {Shahbandeh}, {Skowron}, {Sollerman}, {Soszy{\'n}ski}, {Stritzinger},
  {Szyma{\'n}ski}, {Tartaglia}, {Udalski}, {Ulaczyk}, {Young}, {van Leeuwen},
  \& {van Soelen}}]{Brennan2022c}
{Brennan}, S.~J., {Fraser}, M., {Johansson}, J., {et~al.} 2022{\natexlab{b}},
  \mnras, 513, 5666

\bibitem[{{Brennan} {et~al.}(2023){Brennan}, {Schulze}, {Lunnan}, {Sollerman},
  {Yan}, {Fransson}, {Irani}, {Melinder}, {Chen}, {De}, {Fremling}, {Kim},
  {Perley}, {Pessi}, {Drake}, {Graham}, {Laher}, {Masci}, {Purdum}, \&
  {Rodriguez}}]{Brennan2023}
{Brennan}, S.~J., {Schulze}, S., {Lunnan}, R., {et~al.} 2023, arXiv e-prints,
  arXiv:2312.13280

\bibitem[{{Cardelli} {et~al.}(1989){Cardelli}, {Clayton}, \&
  {Mathis}}]{Cardelli1989}
{Cardelli}, J.~A., {Clayton}, G.~C., \& {Mathis}, J.~S. 1989, \apj, 345, 245

\bibitem[{{Chambers} {et~al.}(2016){Chambers}, {Magnier}, {Metcalfe},
  {Flewelling}, {Huber}, {Waters}, {Denneau}, {Draper}, {Farrow}, {Finkbeiner},
  {Holmberg}, {Koppenhoefer}, {Price}, {Rest}, {Saglia}, {Schlafly}, {Smartt},
  {Sweeney}, {Wainscoat}, {Burgett}, {Chastel}, {Grav}, {Heasley}, {Hodapp},
  {Jedicke}, {Kaiser}, {Kudritzki}, {Luppino}, {Lupton}, {Monet}, {Morgan},
  {Onaka}, {Shiao}, {Stubbs}, {Tonry}, {White}, {Ba{\~n}ados}, {Bell},
  {Bender}, {Bernard}, {Boegner}, {Boffi}, {Botticella}, {Calamida},
  {Casertano}, {Chen}, {Chen}, {Cole}, {Deacon}, {Frenk}, {Fitzsimmons},
  {Gezari}, {Gibbs}, {Goessl}, {Goggia}, {Gourgue}, {Goldman}, {Grant},
  {Grebel}, {Hambly}, {Hasinger}, {Heavens}, {Heckman}, {Henderson}, {Henning},
  {Holman}, {Hopp}, {Ip}, {Isani}, {Jackson}, {Keyes}, {Koekemoer}, {Kotak},
  {Le}, {Liska}, {Long}, {Lucey}, {Liu}, {Martin}, {Masci}, {McLean}, {Mindel},
  {Misra}, {Morganson}, {Murphy}, {Obaika}, {Narayan}, {Nieto-Santisteban},
  {Norberg}, {Peacock}, {Pier}, {Postman}, {Primak}, {Rae}, {Rai}, {Riess},
  {Riffeser}, {Rix}, {R{\"o}ser}, {Russel}, {Rutz}, {Schilbach}, {Schultz},
  {Scolnic}, {Strolger}, {Szalay}, {Seitz}, {Small}, {Smith}, {Soderblom},
  {Taylor}, {Thomson}, {Taylor}, {Thakar}, {Thiel}, {Thilker}, {Unger},
  {Urata}, {Valenti}, {Wagner}, {Walder}, {Walter}, {Watters}, {Werner},
  {Wood-Vasey}, \& {Wyse}}]{Chambers2016}
{Chambers}, K.~C., {Magnier}, E.~A., {Metcalfe}, N., {et~al.} 2016, arXiv
  e-prints, arXiv:1612.05560

\bibitem[{{Chevalier}(2012)}]{Chevalier2012}
{Chevalier}, R.~A. 2012, \apjl, 752, L2

\bibitem[{{Chevalier} \& {Fransson}(2008)}]{Chevalier2008}
{Chevalier}, R.~A. \& {Fransson}, C. 2008, \apjl, 683, L135

\bibitem[{{Coughlin} {et~al.}(2023){Coughlin}, {Bloom}, {Nir}, {Antier}, {du
  Laz}, {van der Walt}, {Crellin-Quick}, {Culino}, {Duev}, {Goldstein},
  {Healy}, {Karambelkar}, {Lilleboe}, {Shin}, {Singer}, {Ahumada}, {Anand},
  {Bellm}, {Dekany}, {Graham}, {Kasliwal}, {Kostadinova}, {Kiendrebeogo},
  {Kulkarni}, {Jenkins}, {LeBaron}, {Mahabal}, {Neill}, {Parazin}, {Peloton},
  {Perley}, {Riddle}, {Rusholme}, {van Santen}, {Sollerman}, {Stein}, {Turpin},
  {Wold}, {Amat}, {Bonnefon}, {Bonnefoy}, {Flament}, {Kerkow}, {Kishore},
  {Jani}, {Mahanty}, {Liu}, {Llinares}, {Makarison}, {Olli{\'e}ric}, {Perez},
  {Pont}, \& {Sharma}}]{Coughlin2023}
{Coughlin}, M.~W., {Bloom}, J.~S., {Nir}, G., {et~al.} 2023, \apjs, 267, 31

\bibitem[{{Crowther}(2012)}]{Crowther2012}
{Crowther}, P. 2012, Astronomy and Geophysics, 53, 4.30

\bibitem[{{Crowther}(2007)}]{Crowther2007}
{Crowther}, P.~A. 2007, \araa, 45, 177

\bibitem[{{Damas-Segovia} {et~al.}(2016){Damas-Segovia}, {Beck}, {Vollmer},
  {Wiegert}, {Krause}, {Irwin}, {We{\.z}gowiec}, {Li}, {Dettmar}, {English}, \&
  {Wang}}]{Damas-Segovia2016}
{Damas-Segovia}, A., {Beck}, R., {Vollmer}, B., {et~al.} 2016, \apj, 824, 30

\bibitem[{{De}(2023)}]{De2023}
{De}, K. 2023, Transient Name Server Discovery Report, 2023-825, 1

\bibitem[{{De} {et~al.}(2020){De}, {Kasliwal}, {Tzanidakis}, {Fremling},
  {Adams}, {Aloisi}, {Andreoni}, {Bagdasaryan}, {Bellm}, {Bildsten},
  {Cannella}, {Cook}, {Delacroix}, {Drake}, {Duev}, {Dugas}, {Frederick},
  {Gal-Yam}, {Goldstein}, {Golkhou}, {Graham}, {Hale}, {Hankins}, {Helou},
  {Ho}, {Irani}, {Jencson}, {Kaplan}, {Kaye}, {Kulkarni}, {Kupfer}, {Laher},
  {Leadbeater}, {Lunnan}, {Masci}, {Miller}, {Neill}, {Ofek}, {Perley},
  {Polin}, {Prince}, {Quataert}, {Reiley}, {Riddle}, {Rusholme}, {Sharma},
  {Shupe}, {Sollerman}, {Tartaglia}, {Walters}, {Yan}, \& {Yao}}]{De2020}
{De}, K., {Kasliwal}, M.~M., {Tzanidakis}, A., {et~al.} 2020, \apj, 905, 58

\bibitem[{{Dekany} {et~al.}(2020){Dekany}, {Smith}, {Riddle}, {Feeney},
  {Porter}, {Hale}, {Zolkower}, {Belicki}, {Kaye}, {Henning}, {Walters},
  {Cromer}, {Delacroix}, {Rodriguez}, {Reiley}, {Mao}, {Hover}, {Murphy},
  {Burruss}, {Baker}, {Kowalski}, {Reif}, {Mueller}, {Bellm}, {Graham}, \&
  {Kulkarni}}]{Dekany2020}
{Dekany}, R., {Smith}, R.~M., {Riddle}, R., {et~al.} 2020, \pasp, 132, 038001

\bibitem[{{Dessart} {et~al.}(2010){Dessart}, {Livne}, \&
  {Waldman}}]{Dessart2010}
{Dessart}, L., {Livne}, E., \& {Waldman}, R. 2010, \mnras, 405, 2113

\bibitem[{{Dong} {et~al.}(2023){Dong}, {Sand}, {Valenti}, {Bostroem},
  {Andrews}, {Hosseinzadeh}, {Hoang}, {Janzen}, {Jencson}, {Lundquist}, {Meza
  Retamal}, {Pearson}, {Shrestha}, {Haislip}, {Kouprianov}, \&
  {Reichart}}]{Dong2023}
{Dong}, Y., {Sand}, D.~J., {Valenti}, S., {et~al.} 2023, \apj, 957, 28

\bibitem[{{Ekholm} {et~al.}(2000){Ekholm}, {Lanoix}, {Teerikorpi},
  {Fouqu{\'e}}, \& {Paturel}}]{Ekholm2000}
{Ekholm}, T., {Lanoix}, P., {Teerikorpi}, P., {Fouqu{\'e}}, P., \& {Paturel},
  G. 2000, \aap, 355, 835

\bibitem[{{Elias-Rosa} {et~al.}(2016){Elias-Rosa}, {Pastorello}, {Benetti},
  {Cappellaro}, {Taubenberger}, {Terreran}, {Fraser}, {Brown}, {Tartaglia},
  {Morales-Garoffolo}, {Harmanen}, {Richardson}, {Artigau}, {Tomasella},
  {Margutti}, {Smartt}, {Dennefeld}, {Turatto}, {Anupama}, {Arbour}, {Berton},
  {Bjorkman}, {Boles}, {Briganti}, {Chornock}, {Ciabattari}, {Cortini},
  {Dimai}, {Gerhartz}, {Itagaki}, {Kotak}, {Mancini}, {Martinelli},
  {Milisavljevic}, {Misra}, {Ochner}, {Patnaude}, {Polshaw}, {Sahu}, \&
  {Zaggia}}]{Elias-Rosa2016}
{Elias-Rosa}, N., {Pastorello}, A., {Benetti}, S., {et~al.} 2016, \mnras, 463,
  3894

\bibitem[{{Foley} {et~al.}(2011){Foley}, {Berger}, {Fox}, {Levesque},
  {Challis}, {Ivans}, {Rhoads}, \& {Soderberg}}]{Foley2011}
{Foley}, R.~J., {Berger}, E., {Fox}, O., {et~al.} 2011, \apj, 732, 32

\bibitem[{{Foley} {et~al.}(2007){Foley}, {Smith}, {Ganeshalingam}, {Li},
  {Chornock}, \& {Filippenko}}]{Foley2007}
{Foley}, R.~J., {Smith}, N., {Ganeshalingam}, M., {et~al.} 2007, \apjl, 657,
  L105

\bibitem[{{Fransson} {et~al.}(2022){Fransson}, {Sollerman}, {Strotjohann},
  {Yang}, {Schulze}, {Barbarino}, {Kool}, {Ofek}, {Crellin-Quick}, {De},
  {Drake}, {Fremling}, {Gal-Yam}, {Ho}, \& {Kasliwal}}]{Fransson2022}
{Fransson}, C., {Sollerman}, J., {Strotjohann}, N.~L., {et~al.} 2022, \aap,
  666, A79

\bibitem[{{Fraser} {et~al.}(2015){Fraser}, {Kotak}, {Pastorello}, {Jerkstrand},
  {Smartt}, {Chen}, {Childress}, {Gilmore}, {Inserra}, {Kankare}, {Margheim},
  {Mattila}, {Valenti}, {Ashall}, {Benetti}, {Botticella}, {Bauer}, {Campbell},
  {Elias-Rosa}, {Fleury}, {Gal-Yam}, {Hachinger}, {Howell}, {Le Guillou},
  {L{\'e}get}, {Morales-Garoffolo}, {Polshaw}, {Spiro}, {Sullivan},
  {Taubenberger}, {Turatto}, {Walker}, {Young}, \& {Zhang}}]{Fraser2015}
{Fraser}, M., {Kotak}, R., {Pastorello}, A., {et~al.} 2015, \mnras, 453, 3886

\bibitem[{{Fraser} {et~al.}(2013){Fraser}, {Magee}, {Kotak}, {Smartt}, {Smith},
  {Polshaw}, {Drake}, {Boles}, {Lee}, {Burgett}, {Chambers}, {Draper},
  {Flewelling}, {Hodapp}, {Kaiser}, {Kudritzki}, {Magnier}, {Price}, {Tonry},
  {Wainscoat}, \& {Waters}}]{Fraser2013}
{Fraser}, M., {Magee}, M., {Kotak}, R., {et~al.} 2013, \apjl, 779, L8

\bibitem[{{Freudling} {et~al.}(2013){Freudling}, {Romaniello}, {Bramich},
  {Ballester}, {Forchi}, {Garc{\'\i}a-Dabl{\'o}}, {Moehler}, \&
  {Neeser}}]{Freudling2013a}
{Freudling}, W., {Romaniello}, M., {Bramich}, D.~M., {et~al.} 2013, \aap, 559,
  A96

\bibitem[{{Fuller} \& {Ro}(2018)}]{Fuller2018}
{Fuller}, J. \& {Ro}, S. 2018, \mnras, 476, 1853

\bibitem[{{Gehrels} {et~al.}(2004){Gehrels}, {Chincarini}, {Giommi}, {Mason},
  {Nousek}, {Wells}, {White}, {Barthelmy}, {Burrows}, {Cominsky}, {Hurley},
  {Marshall}, {M{\'e}sz{\'a}ros}, {Roming}, {Angelini}, {Barbier}, {Belloni},
  {Campana}, {Caraveo}, {Chester}, {Citterio}, {Cline}, {Cropper}, {Cummings},
  {Dean}, {Feigelson}, {Fenimore}, {Frail}, {Fruchter}, {Garmire}, {Gendreau},
  {Ghisellini}, {Greiner}, {Hill}, {Hunsberger}, {Krimm}, {Kulkarni}, {Kumar},
  {Lebrun}, {Lloyd-Ronning}, {Markwardt}, {Mattson}, {Mushotzky}, {Norris},
  {Osborne}, {Paczynski}, {Palmer}, {Park}, {Parsons}, {Paul}, {Rees},
  {Reynolds}, {Rhoads}, {Sasseen}, {Schaefer}, {Short}, {Smale}, {Smith},
  {Stella}, {Tagliaferri}, {Takahashi}, {Tashiro}, {Townsley}, {Tueller},
  {Turner}, {Vietri}, {Voges}, {Ward}, {Willingale}, {Zerbi}, \&
  {Zhang}}]{Gehrels2004}
{Gehrels}, N., {Chincarini}, G., {Giommi}, P., {et~al.} 2004, \apj, 611, 1005

\bibitem[{{Graham} {et~al.}(2019){Graham}, {Kulkarni}, {Bellm}, {Adams},
  {Barbarino}, {Blagorodnova}, {Bodewits}, {Bolin}, {Brady}, {Cenko}, {Chang},
  {Coughlin}, {De}, {Eadie}, {Farnham}, {Feindt}, {Franckowiak}, {Fremling},
  {Gezari}, {Ghosh}, {Goldstein}, {Golkhou}, {Goobar}, {Ho}, {Huppenkothen},
  {Ivezi{\'c}}, {Jones}, {Juric}, {Kaplan}, {Kasliwal}, {Kelley}, {Kupfer},
  {Lee}, {Lin}, {Lunnan}, {Mahabal}, {Miller}, {Ngeow}, {Nugent}, {Ofek},
  {Prince}, {Rauch}, {van Roestel}, {Schulze}, {Singer}, {Sollerman}, {Taddia},
  {Yan}, {Ye}, {Yu}, {Barlow}, {Bauer}, {Beck}, {Belicki}, {Biswas}, {Brinnel},
  {Brooke}, {Bue}, {Bulla}, {Burruss}, {Connolly}, {Cromer}, {Cunningham},
  {Dekany}, {Delacroix}, {Desai}, {Duev}, {Feeney}, {Flynn}, {Frederick},
  {Gal-Yam}, {Giomi}, {Groom}, {Hacopians}, {Hale}, {Helou}, {Henning},
  {Hover}, {Hillenbrand}, {Howell}, {Hung}, {Imel}, {Ip}, {Jackson}, {Kaspi},
  {Kaye}, {Kowalski}, {Kramer}, {Kuhn}, {Landry}, {Laher}, {Mao}, {Masci},
  {Monkewitz}, {Murphy}, {Nordin}, {Patterson}, {Penprase}, {Porter},
  {Rebbapragada}, {Reiley}, {Riddle}, {Rigault}, {Rodriguez}, {Rusholme}, {van
  Santen}, {Shupe}, {Smith}, {Soumagnac}, {Stein}, {Surace}, {Szkody}, {Terek},
  {Van Sistine}, {van Velzen}, {Vestrand}, {Walters}, {Ward}, {Zhang}, \&
  {Zolkower}}]{Graham2019}
{Graham}, M.~J., {Kulkarni}, S.~R., {Bellm}, E.~C., {et~al.} 2019, \pasp, 131,
  078001

\bibitem[{{Heger} {et~al.}(2003){Heger}, {Fryer}, {Woosley}, {Langer}, \&
  {Hartmann}}]{Heger2003}
{Heger}, A., {Fryer}, C.~L., {Woosley}, S.~E., {Langer}, N., \& {Hartmann},
  D.~H. 2003, \apj, 591, 288

\bibitem[{{Hiramatsu} {et~al.}(2023){Hiramatsu}, {Matsumoto}, {Berger},
  {Ransome}, {Villar}, {Gomez}, {Cendes}, {De}, {Farah}, {Howell}, {McCully},
  {Newsome}, {Padilla Gonzalez}, {Pellegrino}, {Suzuki}, \&
  {Terreran}}]{Hiramatsu2023}
{Hiramatsu}, D., {Matsumoto}, T., {Berger}, E., {et~al.} 2023, arXiv e-prints,
  arXiv:2305.11168

\bibitem[{{Hook} {et~al.}(2004){Hook}, {J{\o}rgensen}, {Allington-Smith},
  {Davies}, {Metcalfe}, {Murowinski}, \& {Crampton}}]{Hook2004}
{Hook}, I.~M., {J{\o}rgensen}, I., {Allington-Smith}, J.~R., {et~al.} 2004,
  \pasp, 116, 425

\bibitem[{{Hosseinzadeh} {et~al.}(2017){Hosseinzadeh}, {Arcavi}, {Valenti},
  {McCully}, {Howell}, {Johansson}, {Sollerman}, {Pastorello}, {Benetti},
  {Cao}, {Cenko}, {Clubb}, {Corsi}, {Duggan}, {Elias-Rosa}, {Filippenko},
  {Fox}, {Fremling}, {Horesh}, {Karamehmetoglu}, {Kasliwal}, {Marion}, {Ofek},
  {Sand}, {Taddia}, {Zheng}, {Fraser}, {Gal-Yam}, {Inserra}, {Laher}, {Masci},
  {Rebbapragada}, {Smartt}, {Smith}, {Sullivan}, {Surace}, \&
  {Wo{\'z}niak}}]{Hosseinzadeh2017}
{Hosseinzadeh}, G., {Arcavi}, I., {Valenti}, S., {et~al.} 2017, \apj, 836, 158

\bibitem[{{Hosseinzadeh} {et~al.}(2022){Hosseinzadeh}, {Kilpatrick}, {Dong},
  {Sand}, {Andrews}, {Bostroem}, {Janzen}, {Jencson}, {Lundquist}, {Meza
  Retamal}, {Pearson}, {Valenti}, {Wyatt}, {Burke}, {Hiramatsu}, {Howell},
  {McCully}, {Newsome}, {Gonzalez}, {Pellegrino}, {Terreran}, {Auchettl},
  {Davis}, {Foley}, {Miao}, {Pan}, {Rest}, {Siebert}, {Taggart}, {Tucker},
  {Cyrus Leung}, {Swift}, {Yang}, {Anderson}, {Ashall}, {Benetti}, {Brown},
  {Cartier}, {Chen}, {Della Valle}, {Galbany}, {Gomez}, {Gromadzki}, {Haislip},
  {Hsiao}, {Inserra}, {Jha}, {Killestein}, {Kouprianov}, {Kozyreva},
  {M{\"u}ller-Bravo}, {Nicholl}, {Paraskeva}, {Reichart}, {Ryder},
  {Shahbandeh}, {Shappee}, {Smith}, \& {Young}}]{Hosseinzadeh2022}
{Hosseinzadeh}, G., {Kilpatrick}, C.~D., {Dong}, Y., {et~al.} 2022, \apj, 935,
  31

\bibitem[{{Humphreys} \& {Davidson}(1994)}]{Humphreys1994}
{Humphreys}, R.~M. \& {Davidson}, K. 1994, \pasp, 106, 1025

\bibitem[{{Irani} {et~al.}(2023){Irani}, {Morag}, {Gal-Yam}, {Waxman},
  {Schulze}, {Sollerman}, {Hinds}, {Perley}, {Chen}, {Strotjohann}, {Yaron},
  {Zimmerman}, {Bruch}, {Ofek}, {Soumagnac}, {Yang}, {Groom}, {Masci},
  {Riddle}, {Bellm}, \& {Hale}}]{Irani2023}
{Irani}, I., {Morag}, J., {Gal-Yam}, A., {et~al.} 2023, arXiv e-prints,
  arXiv:2310.16885

\bibitem[{{Ivanova} {et~al.}(2013){Ivanova}, {Justham}, {Chen}, {De Marco},
  {Fryer}, {Gaburov}, {Ge}, {Glebbeek}, {Han}, {Li}, {Lu}, {Marsh},
  {Podsiadlowski}, {Potter}, {Soker}, {Taam}, {Tauris}, {van den Heuvel}, \&
  {Webbink}}]{Ivanova2013}
{Ivanova}, N., {Justham}, S., {Chen}, X., {et~al.} 2013, \aapr, 21, 59

\bibitem[{{Jacobson-Gal{\'a}n} {et~al.}(2022){Jacobson-Gal{\'a}n}, {Dessart},
  {Jones}, {Margutti}, {Coppejans}, {Dimitriadis}, {Foley}, {Kilpatrick},
  {Matthews}, {Rest}, {Terreran}, {Aleo}, {Auchettl}, {Blanchard}, {Coulter},
  {Davis}, {de Boer}, {DeMarchi}, {Drout}, {Earl}, {Gagliano}, {Gall},
  {Hjorth}, {Huber}, {Ibik}, {Milisavljevic}, {Pan}, {Rest}, {Ridden-Harper},
  {Rojas-Bravo}, {Siebert}, {Smith}, {Taggart}, {Tinyanont}, {Wang}, \&
  {Zenati}}]{Jacobson2022}
{Jacobson-Gal{\'a}n}, W.~V., {Dessart}, L., {Jones}, D.~O., {et~al.} 2022,
  \apj, 924, 15

\bibitem[{{Janka}(2012)}]{Janka2012}
{Janka}, H.-T. 2012, Annual Review of Nuclear and Particle Science, 62, 407

\bibitem[{{Jerkstrand} {et~al.}(2014){Jerkstrand}, {Smartt}, {Fraser},
  {Fransson}, {Sollerman}, {Taddia}, \& {Kotak}}]{Jerkstrand2014}
{Jerkstrand}, A., {Smartt}, S.~J., {Fraser}, M., {et~al.} 2014, \mnras, 439,
  3694

\bibitem[{{Karamehmetoglu} {et~al.}(2021){Karamehmetoglu}, {Fransson},
  {Sollerman}, {Tartaglia}, {Taddia}, {De}, {Fremling}, {Bagdasaryan},
  {Barbarino}, {Bellm}, {Dekany}, {Dugas}, {Giomi}, {Goobar}, {Graham}, {Ho},
  {Laher}, {Masci}, {Neill}, {Perley}, {Riddle}, {Rusholme}, \&
  {Soumagnac}}]{Karamehmetoglu2021}
{Karamehmetoglu}, E., {Fransson}, C., {Sollerman}, J., {et~al.} 2021, \aap,
  649, A163

\bibitem[{{Kashi} {et~al.}(2013){Kashi}, {Soker}, \& {Moskovitz}}]{Kashi2013}
{Kashi}, A., {Soker}, N., \& {Moskovitz}, N. 2013, \mnras, 436, 2484

\bibitem[{{Khatami} \& {Kasen}(2023)}]{Khatami2023}
{Khatami}, D. \& {Kasen}, D. 2023, arXiv e-prints, arXiv:2304.03360

\bibitem[{{Khatami} \& {Kasen}(2019)}]{Khatami2019}
{Khatami}, D.~K. \& {Kasen}, D.~N. 2019, \apj, 878, 56

\bibitem[{{Kilpatrick} {et~al.}(2018){Kilpatrick}, {Foley}, {Drout}, {Pan},
  {Panther}, {Coulter}, {Filippenko}, {Marion}, {Piro}, {Rest}, {Seitenzahl},
  {Strampelli}, \& {Wang}}]{Kilpatrick2018}
{Kilpatrick}, C.~D., {Foley}, R.~J., {Drout}, M.~R., {et~al.} 2018, \mnras,
  473, 4805

\bibitem[{{Kim} {et~al.}(2022){Kim}, {Rigault}, {Neill}, {Briday}, {Copin},
  {Lezmy}, {Nicolas}, {Riddle}, {Sharma}, {Smith}, {Sollerman}, \&
  {Walters}}]{Kim2022}
{Kim}, Y.~L., {Rigault}, M., {Neill}, J.~D., {et~al.} 2022, \pasp, 134, 024505

\bibitem[{{Kuncarayakti} {et~al.}(2015){Kuncarayakti}, {Maeda}, {Bersten},
  {Folatelli}, {Morrell}, {Hsiao}, {Gonz{\'a}lez-Gait{\'a}n}, {Anderson},
  {Hamuy}, {de Jaeger}, {Guti{\'e}rrez}, \& {Kawabata}}]{Kuncarayakti2015}
{Kuncarayakti}, H., {Maeda}, K., {Bersten}, M.~C., {et~al.} 2015, \aap, 579,
  A95

\bibitem[{{Labrie} {et~al.}(2019){Labrie}, {Anderson}, {C{\'a}rdenes},
  {Simpson}, \& {Turner}}]{Labrie2019}
{Labrie}, K., {Anderson}, K., {C{\'a}rdenes}, R., {Simpson}, C., \& {Turner},
  J. E.~H. 2019, in Astronomical Society of the Pacific Conference Series, Vol.
  523, Astronomical Data Analysis Software and Systems XXVII, ed. P.~J.
  {Teuben}, M.~W. {Pound}, B.~A. {Thomas}, \& E.~M. {Warner}, 321

\bibitem[{{Leloudas} {et~al.}(2015){Leloudas}, {Hsiao}, {Johansson}, {Maeda},
  {Moriya}, {Nordin}, {Petrushevska}, {Silverman}, {Sollerman}, {Stritzinger},
  {Taddia}, \& {Xu}}]{Leloudas2015}
{Leloudas}, G., {Hsiao}, E.~Y., {Johansson}, J., {et~al.} 2015, \aap, 574, A61

\bibitem[{{Leung} {et~al.}(2021){Leung}, {Wu}, \& {Fuller}}]{Leung2021}
{Leung}, S.-C., {Wu}, S., \& {Fuller}, J. 2021, \apj, 923, 41

\bibitem[{{Lianou} {et~al.}(2019){Lianou}, {Barmby}, {Mosenkov}, {Lehnert}, \&
  {Karczewski}}]{Lianou2019}
{Lianou}, S., {Barmby}, P., {Mosenkov}, A.~A., {Lehnert}, M., \& {Karczewski},
  O. 2019, \aap, 631, A38

\bibitem[{{Mandigo-Stoba} {et~al.}(2022){Mandigo-Stoba}, {Fremling}, \&
  {Kasliwal}}]{Mandigo2022}
{Mandigo-Stoba}, M.~S., {Fremling}, C., \& {Kasliwal}, M. 2022, The Journal of
  Open Source Software, 7, 3612

\bibitem[{{Margutti} {et~al.}(2014){Margutti}, {Milisavljevic}, {Soderberg},
  {Chornock}, {Zauderer}, {Murase}, {Guidorzi}, {Sanders}, {Kuin}, {Fransson},
  {Levesque}, {Chandra}, {Berger}, {Bianco}, {Brown}, {Challis},
  {Chatzopoulos}, {Cheung}, {Choi}, {Chomiuk}, {Chugai}, {Contreras}, {Drout},
  {Fesen}, {Foley}, {Fong}, {Friedman}, {Gall}, {Gehrels}, {Hjorth}, {Hsiao},
  {Kirshner}, {Im}, {Leloudas}, {Lunnan}, {Marion}, {Martin}, {Morrell},
  {Neugent}, {Omodei}, {Phillips}, {Rest}, {Silverman}, {Strader},
  {Stritzinger}, {Szalai}, {Utterback}, {Vinko}, {Wheeler}, {Arnett},
  {Campana}, {Chevalier}, {Ginsburg}, {Kamble}, {Roming}, {Pritchard}, \&
  {Stringfellow}}]{Margutti2014}
{Margutti}, R., {Milisavljevic}, D., {Soderberg}, A.~M., {et~al.} 2014, \apj,
  780, 21

\bibitem[{{Masci} {et~al.}(2019){Masci}, {Laher}, {Rusholme}, {Shupe}, {Groom},
  {Surace}, {Jackson}, {Monkewitz}, {Beck}, {Flynn}, {Terek}, {Landry},
  {Hacopians}, {Desai}, {Howell}, {Brooke}, {Imel}, {Wachter}, {Ye}, {Lin},
  {Cenko}, {Cunningham}, {Rebbapragada}, {Bue}, {Miller}, {Mahabal}, {Bellm},
  {Patterson}, {Juri{\'c}}, {Golkhou}, {Ofek}, {Walters}, {Graham}, {Kasliwal},
  {Dekany}, {Kupfer}, {Burdge}, {Cannella}, {Barlow}, {Van Sistine}, {Giomi},
  {Fremling}, {Blagorodnova}, {Levitan}, {Riddle}, {Smith}, {Helou}, {Prince},
  \& {Kulkarni}}]{Masci2019}
{Masci}, F.~J., {Laher}, R.~R., {Rusholme}, B., {et~al.} 2019, \pasp, 131,
  018003

\bibitem[{{Matsumoto} \& {Metzger}(2022)}]{Matsumoto2022}
{Matsumoto}, T. \& {Metzger}, B.~D. 2022, \apj, 936, 114

\bibitem[{{Mattila} {et~al.}(2008){Mattila}, {Meikle}, {Lundqvist},
  {Pastorello}, {Kotak}, {Eldridge}, {Smartt}, {Adamson}, {Gerardy}, {Rizzi},
  {Stephens}, \& {van Dyk}}]{Mattila2008}
{Mattila}, S., {Meikle}, W.~P.~S., {Lundqvist}, P., {et~al.} 2008, \mnras, 389,
  141

\bibitem[{{Mauerhan} {et~al.}(2013){Mauerhan}, {Smith}, {Filippenko},
  {Blanchard}, {Blanchard}, {Casper}, {Cenko}, {Clubb}, {Cohen}, {Fuller},
  {Li}, \& {Silverman}}]{Mauerhan2013}
{Mauerhan}, J.~C., {Smith}, N., {Filippenko}, A.~V., {et~al.} 2013, \mnras,
  430, 1801

\bibitem[{{Modjaz} {et~al.}(2009){Modjaz}, {Li}, {Butler}, {Chornock},
  {Perley}, {Blondin}, {Bloom}, {Filippenko}, {Kirshner}, {Kocevski},
  {Poznanski}, {Hicken}, {Foley}, {Stringfellow}, {Berlind}, {Barrado y
  Navascues}, {Blake}, {Bouy}, {Brown}, {Challis}, {Chen}, {de Vries},
  {Dufour}, {Falco}, {Friedman}, {Ganeshalingam}, {Garnavich}, {Holden},
  {Illingworth}, {Lee}, {Liebert}, {Marion}, {Olivier}, {Prochaska},
  {Silverman}, {Smith}, {Starr}, {Steele}, {Stockton}, {Williams}, \&
  {Wood-Vasey}}]{Modjaz2009}
{Modjaz}, M., {Li}, W., {Butler}, N., {et~al.} 2009, \apj, 702, 226

\bibitem[{{M{\"u}ller} {et~al.}(2016){M{\"u}ller}, {SViallet}, {Heger}, \&
  {Janka}}]{Muller2016}
{M{\"u}ller}, B., {SViallet}, M., {Heger}, A., \& {Janka}, H.-T. 2016, \apj,
  833, 124

\bibitem[{{Nyholm} {et~al.}(2017){Nyholm}, {Sollerman}, {Taddia}, {Fremling},
  {Moriya}, {Ofek}, {Gal-Yam}, {De Cia}, {Roy}, {Kasliwal}, {Cao}, {Nugent}, \&
  {Masci}}]{Nyholm2017}
{Nyholm}, A., {Sollerman}, J., {Taddia}, F., {et~al.} 2017, \aap, 605, A6

\bibitem[{{Ofek} {et~al.}(2013){Ofek}, {Sullivan}, {Cenko}, {Kasliwal},
  {Gal-Yam}, {Kulkarni}, {Arcavi}, {Bildsten}, {Bloom}, {Horesh}, {Howell},
  {Filippenko}, {Laher}, {Murray}, {Nakar}, {Nugent}, {Silverman}, {Shaviv},
  {Surace}, \& {Yaron}}]{Ofek2013}
{Ofek}, E.~O., {Sullivan}, M., {Cenko}, S.~B., {et~al.} 2013, \nat, 494, 65

\bibitem[{{Ofek} {et~al.}(2014){Ofek}, {Sullivan}, {Shaviv}, {Steinbok},
  {Arcavi}, {Gal-Yam}, {Tal}, {Kulkarni}, {Nugent}, {Ben-Ami}, {Kasliwal},
  {Cenko}, {Laher}, {Surace}, {Bloom}, {Filippenko}, {Silverman}, \&
  {Yaron}}]{Ofe2014}
{Ofek}, E.~O., {Sullivan}, M., {Shaviv}, N.~J., {et~al.} 2014, \apj, 789, 104

\bibitem[{{Oke} {et~al.}(1995){Oke}, {Cohen}, {Carr}, {Cromer}, {Dingizian},
  {Harris}, {Labrecque}, {Lucinio}, {Schaal}, {Epps}, \& {Miller}}]{Oke1995}
{Oke}, J.~B., {Cohen}, J.~G., {Carr}, M., {et~al.} 1995, \pasp, 107, 375

\bibitem[{{Osterbrock}(1989)}]{Osterbrock1989}
{Osterbrock}, D.~E. 1989, {Astrophysics of gaseous nebulae and active galactic
  nuclei} (University Science Books)

\bibitem[{{Pastorello} {et~al.}(2013){Pastorello}, {Cappellaro}, {Inserra},
  {Smartt}, {Pignata}, {Benetti}, {Valenti}, {Fraser}, {Tak{\'a}ts}, {Benitez},
  {Botticella}, {Brimacombe}, {Bufano}, {Cellier-Holzem}, {Costado}, {Cupani},
  {Curtis}, {Elias-Rosa}, {Ergon}, {Fynbo}, {Hambsch}, {Hamuy}, {Harutyunyan},
  {Ivarson}, {Kankare}, {Martin}, {Kotak}, {LaCluyze}, {Maguire}, {Mattila},
  {Maza}, {McCrum}, {Miluzio}, {Norgaard-Nielsen}, {Nysewander}, {Ochner},
  {Pan}, {Pumo}, {Reichart}, {Tan}, {Taubenberger}, {Tomasella}, {Turatto}, \&
  {Wright}}]{Pastorello2013}
{Pastorello}, A., {Cappellaro}, E., {Inserra}, C., {et~al.} 2013, \apj, 767, 1

\bibitem[{{Pastorello} {et~al.}(2007){Pastorello}, {Smartt}, {Mattila},
  {Eldridge}, {Young}, {Itagaki}, {Yamaoka}, {Navasardyan}, {Valenti}, {Patat},
  {Agnoletto}, {Augusteijn}, {Benetti}, {Cappellaro}, {Boles}, {Bonnet-Bidaud},
  {Botticella}, {Bufano}, {Cao}, {Deng}, {Dennefeld}, {Elias-Rosa},
  {Harutyunyan}, {Keenan}, {Iijima}, {Lorenzi}, {Mazzali}, {Meng}, {Nakano},
  {Nielsen}, {Smoker}, {Stanishev}, {Turatto}, {Xu}, \&
  {Zampieri}}]{Pastorello2007}
{Pastorello}, A., {Smartt}, S.~J., {Mattila}, S., {et~al.} 2007, \nat, 447, 829

\bibitem[{{Perley}(2019)}]{Perley2019}
{Perley}, D.~A. 2019, \pasp, 131, 084503

\bibitem[{{Piro}(2015)}]{Piro2015}
{Piro}, A.~L. 2015, \apjl, 808, L51

\bibitem[{{Prieto} {et~al.}(2014){Prieto}, {Rest}, {Bianco}, {Matheson},
  {Smith}, {Walborn}, {Hsiao}, {Chornock}, {Paredes {\'A}lvarez}, {Campillay},
  {Contreras}, {Gonz{\'a}lez}, {James}, {Knapp}, {Kunder}, {Margheim},
  {Morrell}, {Phillips}, {Smith}, {Welch}, \& {Zenteno}}]{Prieto2014}
{Prieto}, J.~L., {Rest}, A., {Bianco}, F.~B., {et~al.} 2014, \apjl, 787, L8

\bibitem[{{Prochaska} {et~al.}(2020{\natexlab{a}}){Prochaska}, {Hennawi},
  {Cooke}, {Westfall}, {Wang}, {EmAstro}, {Tiffanyhsyu}, {Wasserman},
  {Villaume}, {Marijana777}, {Schindler}, {Young}, {Simha}, {Wilde}, {Tejos},
  {Isbell}, {Fl{\"o}rs}, {Sandford}, {Vasovi{\'c}}, {Betts}, \&
  {Holden}}]{pypeit:zenodo}
{Prochaska}, J.~X., {Hennawi}, J., {Cooke}, R., {et~al.} 2020{\natexlab{a}},
  {pypeit/PypeIt: Release 1.0.0}

\bibitem[{{Prochaska} {et~al.}(2020{\natexlab{b}}){Prochaska}, {Hennawi},
  {Westfall}, {Cooke}, {Wang}, {Hsyu}, {Davies}, \&
  {Farina}}]{pypeit:joss_arXiv}
{Prochaska}, J.~X., {Hennawi}, J.~F., {Westfall}, K.~B., {et~al.}
  2020{\natexlab{b}}, arXiv e-prints, arXiv:2005.06505

\bibitem[{Prochaska {et~al.}(2020)Prochaska, Hennawi, Westfall, Cooke, Wang,
  Hsyu, Davies, Farina, \& Pelliccia}]{pypeit:joss_pub}
Prochaska, J.~X., Hennawi, J.~F., Westfall, K.~B., {et~al.} 2020, Journal of
  Open Source Software, 5, 2308

\bibitem[{{Pursiainen} {et~al.}(2022){Pursiainen}, {Leloudas}, {Paraskeva},
  {Cikota}, {Anderson}, {Angus}, {Brennan}, {Bulla}, {Camacho-I{\~n}iguez},
  {Charalampopoulos}, {Chen}, {Delgado Manche{\~n}o}, {Fraser}, {Frohmaier},
  {Galbany}, {Guti{\'e}rrez}, {Gromadzki}, {Inserra}, {Maund},
  {M{\"u}ller-Bravo}, {Mu{\~n}oz Torres}, {Nicholl}, {Onori}, {Patat}, {Pessi},
  {Roy}, {Spyromilio}, {Wiseman}, \& {Young}}]{Pursiainen2022}
{Pursiainen}, M., {Leloudas}, G., {Paraskeva}, E., {et~al.} 2022, \aap, 666,
  A30

\bibitem[{{Quataert} \& {Shiode}(2012)}]{Quataert2012}
{Quataert}, E. \& {Shiode}, J. 2012, \mnras, 423, L92

\bibitem[{{Ransome} {et~al.}(2023){Ransome}, {Villar}, {Tartaglia}, {Gonzalez},
  {Jacobson-Gal{\'a}n}, {Kilpatrick}, {Margutti}, {Foley}, {Grayling}, {Ni},
  {Yarza}, {Ye}, {Auchettl}, {de Boer}, {Chambers}, {Coulter}, {Drout},
  {Farias}, {Gall}, {Gao}, {Huber}, {Ibik}, {Jones}, {Khetan}, {Lin},
  {Politsch}, {Raimundo}, {Rest}, {Wainscoat}, {Karthik Yadavalli}, \&
  {Zenati}}]{Ransome2023}
{Ransome}, C.~L., {Villar}, V.~A., {Tartaglia}, A., {et~al.} 2023, arXiv
  e-prints, arXiv:2312.04426

\bibitem[{{Rigault} {et~al.}(2019){Rigault}, {Neill}, {Blagorodnova}, {Dugas},
  {Feeney}, {Walters}, {Brinnel}, {Copin}, {Fremling}, {Nordin}, \&
  {Sollerman}}]{Rigault2019}
{Rigault}, M., {Neill}, J.~D., {Blagorodnova}, N., {et~al.} 2019, \aap, 627,
  A115

\bibitem[{{Roming} {et~al.}(2005){Roming}, {Kennedy}, {Mason}, {Nousek}, {Ahr},
  {Bingham}, {Broos}, {Carter}, {Hancock}, {Huckle}, {Hunsberger}, {Kawakami},
  {Killough}, {Koch}, {McLelland}, {Smith}, {Smith}, {Soto}, {Boyd},
  {Breeveld}, {Holland}, {Ivanushkina}, {Pryzby}, {Still}, \&
  {Stock}}]{Roming2005}
{Roming}, P. W.~A., {Kennedy}, T.~E., {Mason}, K.~O., {et~al.} 2005, \ssr, 120,
  95

\bibitem[{{Schlafly} \& {Finkbeiner}(2011)}]{Schlafly2011}
{Schlafly}, E.~F. \& {Finkbeiner}, D.~P. 2011, \apj, 737, 103

\bibitem[{{Shingles} {et~al.}(2021){Shingles}, {Smith}, {Young}, {Smartt},
  {Tonry}, {Denneau}, {Heinze}, {Weiland}, {Flewelling}, {Stalder},
  {Clocchiatti}, {F{\"o}rster}, {Pignata}, {Rest}, {Anderson}, {Stubbs}, \&
  {Erasmus}}]{Shingles2021a}
{Shingles}, L., {Smith}, K.~W., {Young}, D.~R., {et~al.} 2021, Transient Name
  Server AstroNote, 7, 1

\bibitem[{{Siebert} {et~al.}(2019){Siebert}, {Foley}, {Jones}, {Angulo},
  {Davis}, {Duarte}, {Strasburger}, {Conlon}, {Kazmi}, {Nishimoto}, {Schubert},
  {Sun}, \& {Tippens}}]{Siebert2019}
{Siebert}, M.~R., {Foley}, R.~J., {Jones}, D.~O., {et~al.} 2019, \mnras, 486,
  5785

\bibitem[{{Smith} {et~al.}(2020){Smith}, {Smartt}, {Young}, {Tonry}, {Denneau},
  {Flewelling}, {Heinze}, {Weiland}, {Stalder}, {Rest}, {Stubbs}, {Anderson},
  {Chen}, {Clark}, {Do}, {F{\"o}rster}, {Fulton}, {Gillanders}, {McBrien},
  {O'Neill}, {Srivastav}, \& {Wright}}]{Smith2020}
{Smith}, K.~W., {Smartt}, S.~J., {Young}, D.~R., {et~al.} 2020, \pasp, 132,
  085002

\bibitem[{{Smith}(2011)}]{Smith2011}
{Smith}, N. 2011, \mnras, 415, 2020

\bibitem[{{Smith} {et~al.}(2022){Smith}, {Andrews}, {Filippenko}, {Fox},
  {Mauerhan}, \& {Van Dyk}}]{Smith2022}
{Smith}, N., {Andrews}, J.~E., {Filippenko}, A.~V., {et~al.} 2022, \mnras, 515,
  71

\bibitem[{{Smith} \& {Arnett}(2014)}]{Smith2014a}
{Smith}, N. \& {Arnett}, W.~D. 2014, \apj, 785, 82

\bibitem[{{Smith} {et~al.}(2010){Smith}, {Miller}, {Li}, {Filippenko},
  {Silverman}, {Howard}, {Nugent}, {Marcy}, {Bloom}, {Ghez}, {Lu}, {Yelda},
  {Bernstein}, \& {Colucci}}]{Smith2010}
{Smith}, N., {Miller}, A., {Li}, W., {et~al.} 2010, \aj, 139, 1451

\bibitem[{{Smith} {et~al.}(2023){Smith}, {Pearson}, {Sand}, {Ilyin},
  {Bostroem}, {Hosseinzadeh}, \& {Shrestha}}]{Smith2023a}
{Smith}, N., {Pearson}, J., {Sand}, D.~J., {et~al.} 2023, \apj, 956, 46

\bibitem[{{Smith} {et~al.}(2018){Smith}, {Rest}, {Andrews}, {Matheson},
  {Bianco}, {Prieto}, {James}, {Smith}, {Strampelli}, \& {Zenteno}}]{Smith2018}
{Smith}, N., {Rest}, A., {Andrews}, J.~E., {et~al.} 2018, \mnras, 480, 1457

\bibitem[{{Strotjohann} {et~al.}(2021){Strotjohann}, {Ofek}, {Gal-Yam},
  {Bruch}, {Schulze}, {Shaviv}, {Sollerman}, {Filippenko}, {Yaron}, {Fremling},
  {Nordin}, {Kool}, {Perley}, {Ho}, {Yang}, {Yao}, {Soumagnac}, {Graham},
  {Barbarino}, {Tartaglia}, {De}, {Goldstein}, {Cook}, {Brink}, {Taggart},
  {Yan}, {Lunnan}, {Kasliwal}, {Kulkarni}, {Nugent}, {Masci}, {Rosnet},
  {Adams}, {Andreoni}, {Bagdasaryan}, {Bellm}, {Burdge}, {Duev}, {Dugas},
  {Frederick}, {Goldwasser}, {Hankins}, {Irani}, {Karambelkar}, {Kupfer},
  {Liang}, {Neill}, {Porter}, {Riddle}, {Sharma}, {Short}, {Taddia},
  {Tzanidakis}, {van Roestel}, {Walters}, \& {Zhuang}}]{Strotjohann2021}
{Strotjohann}, N.~L., {Ofek}, E.~O., {Gal-Yam}, A., {et~al.} 2021, \apj, 907,
  99

\bibitem[{{Th{\"o}ne} {et~al.}(2017){Th{\"o}ne}, {de Ugarte Postigo},
  {Leloudas}, {Gall}, {Cano}, {Maeda}, {Schulze}, {Campana}, {Wiersema},
  {Groh}, {de la Rosa}, {Bauer}, {Malesani}, {Maund}, {Morrell}, \&
  {Beletsky}}]{Thone2017}
{Th{\"o}ne}, C.~C., {de Ugarte Postigo}, A., {Leloudas}, G., {et~al.} 2017,
  \aap, 599, A129

\bibitem[{{Tonry} {et~al.}(2018){Tonry}, {Denneau}, {Heinze}, {Stalder},
  {Smith}, {Smartt}, {Stubbs}, {Weiland}, \& {Rest}}]{Tonry2018}
{Tonry}, J.~L., {Denneau}, L., {Heinze}, A.~N., {et~al.} 2018, \pasp, 130,
  064505

\bibitem[{{Tsang} {et~al.}(2022){Tsang}, {Kasen}, \& {Bildsten}}]{Tsang2022}
{Tsang}, B. T.~H., {Kasen}, D., \& {Bildsten}, L. 2022, \apj, 936, 28

\bibitem[{{Tsuna} {et~al.}(2024){Tsuna}, {Matsumoto}, {Wu}, \&
  {Fuller}}]{Tsuna2024}
{Tsuna}, D., {Matsumoto}, T., {Wu}, S.~C., \& {Fuller}, J. 2024, arXiv
  e-prints, arXiv:2401.02389

\bibitem[{{Valerin} {et~al.}(2023){Valerin}, {Benetti},
  {Elias{\textendash}Rosa}, {Korn}, {Liimets}, {Pastorello}, {Reguitti},
  {Tartaglia}, {Kankare}, {Fraser}, {Stritzinger}, {Lundqvist}, {Tomasella},
  {Cappellaro}, \& {Salmaso}}]{Valerin2023}
{Valerin}, G., {Benetti}, S., {Elias{\textendash}Rosa}, N., {et~al.} 2023,
  Transient Name Server Classification Report, 2023-1777, 1

\bibitem[{{van der Walt} {et~al.}(2019){van der Walt}, {Crellin-Quick}, \&
  {Bloom}}]{Walt2019}
{van der Walt}, S., {Crellin-Quick}, A., \& {Bloom}, J. 2019, The Journal of
  Open Source Software, 4, 1247

\bibitem[{{Van Dyk} {et~al.}(2000){Van Dyk}, {Peng}, {King}, {Filippenko},
  {Treffers}, {Li}, \& {Richmond}}]{VanDyk2000}
{Van Dyk}, S.~D., {Peng}, C.~Y., {King}, J.~Y., {et~al.} 2000, \pasp, 112, 1532

\bibitem[{{Wang} {et~al.}(2022){Wang}, {Liu}, {Lin}, {Wang}, {Dai}, {Li}, \&
  {Song}}]{Wang2022}
{Wang}, L.-J., {Liu}, L.-D., {Lin}, W.-L., {et~al.} 2022, \apj, 933, 102

\bibitem[{{Weilbacher} {et~al.}(2020){Weilbacher}, {Palsa}, {Streicher},
  {Bacon}, {Urrutia}, {Wisotzki}, {Conseil}, {Husemann}, {Jarno}, {Kelz},
  {P{\'e}contal-Rousset}, {Richard}, {Roth}, {Selman}, \&
  {Vernet}}]{Weilbacher2020a}
{Weilbacher}, P.~M., {Palsa}, R., {Streicher}, O., {et~al.} 2020, \aap, 641,
  A28

\bibitem[{{Woosley}(2018)}]{Woosley2018}
{Woosley}, S.~E. 2018, \apj, 863, 105

\bibitem[{{Woosley} {et~al.}(2007){Woosley}, {Blinnikov}, \&
  {Heger}}]{Woosley2007}
{Woosley}, S.~E., {Blinnikov}, S., \& {Heger}, A. 2007, \nat, 450, 390

\bibitem[{{Woosley} \& {Weaver}(1995)}]{Woosley1995}
{Woosley}, S.~E. \& {Weaver}, T.~A. 1995, \apjs, 101, 181

\bibitem[{{Wu} \& {Fuller}(2022)}]{Wu2022}
{Wu}, S. \& {Fuller}, J. 2022, in American Astronomical Society Meeting
  Abstracts, Vol.~54, American Astronomical Society Meeting Abstracts, 327.07

\bibitem[{{Yaron} \& {Gal-Yam}(2012)}]{Yaron2012}
{Yaron}, O. \& {Gal-Yam}, A. 2012, \pasp, 124, 668

\bibitem[{{Yoon} {et~al.}(2017){Yoon}, {Dessart}, \& {Clocchiatti}}]{Yoon2017}
{Yoon}, S.-C., {Dessart}, L., \& {Clocchiatti}, A. 2017, \apj, 840, 10

\bibitem[{{Yoon} {et~al.}(2010){Yoon}, {Woosley}, \& {Langer}}]{Yoon2010}
{Yoon}, S.~C., {Woosley}, S.~E., \& {Langer}, N. 2010, \apj, 725, 940

\bibitem[{{Zapartas} {et~al.}(2021){Zapartas}, {de Mink}, {Justham}, {Smith},
  {Renzo}, \& {de Koter}}]{Zapartas2021}
{Zapartas}, E., {de Mink}, S.~E., {Justham}, S., {et~al.} 2021, \aap, 645, A6

\bibitem[{{Zimmerman} {et~al.}(2023){Zimmerman}, {Irani}, {Chen}, {Gal-Yam},
  {Schulze}, {Perley}, {Sollerman}, {Filippenko}, {Shenar}, {Yaron}, {Shahaf},
  {Bruch}, {Ofek}, {De Cia}, {Brink}, {Yang}, {Vasylyev}, {Ben Ami}, {Aubert},
  {Badash}, {Bloom}, {Brown}, {De}, {Dimitriadis}, {Fransson}, {Fremling},
  {Hinds}, {Horesh}, {Johansson}, {Kasliwal}, {Kulkarni}, {Kushnir}, {Martin},
  {Matuzewski}, {McGurk}, {Miller}, {Morag}, {Neil}, {Nugent}, {Post},
  {Prusinski}, {Qin}, {Raichoor}, {Riddle}, {Rowe}, {Rusholme}, {Sfaradi},
  {Sjoberg}, {Soumagnac}, {Stein}, {Strotjohann}, {Terwel}, {Wasserman},
  {Wise}, {Wold}, {Yan}, \& {Zhang}}]{Zimmerman2023}
{Zimmerman}, E.~A., {Irani}, I., {Chen}, P., {et~al.} 2023, arXiv e-prints,
  arXiv:2310.10727

\end{thebibliography}

\end{document}